\documentclass[noshowpacs,prl,onecolumn,preprintnumbers,superscriptaddress,amsmath,amssymb,floatfix]{revtex4}

\usepackage[utf8]{inputenc}

\usepackage{enumitem}
\usepackage{graphicx}
\graphicspath{{images/}}
\usepackage[caption=false,justification=raggedright, size=small]{subfig}
\usepackage{placeins}
\usepackage{mathtools}
\usepackage{amsthm}
\usepackage{amsmath}
\usepackage{color,soul}
\usepackage{xkeyval,xcolor}
\usepackage{algcompatible}
\usepackage{newfloat}
\usepackage{comment}
\usepackage{ragged2e}
\usepackage{placeins}
\usepackage{fancyhdr}
\usepackage{lipsum, babel}
\usepackage{verbatim}
\usepackage{setspace}

\usepackage[super]{nth}

\begin{document}
\title{Robustness on distributed coupling networks with multiple dependent links from finite functional components}
\author{Gaogao Dong, Nannan Sun, Fan Wang}
\affiliation{School of Mathematical Sciences, Jiangsu University, Zhenjiang, 212013 Jiangsu, China}

\setstretch{1.3} 
 \begin{abstract}
 \vspace{1cm}
 \section*{Abstract}
 \fontsize{11}{16}\selectfont
      The rapid advancement of technology underscores the critical importance of robustness in complex network systems. This paper presents a framework for investigating the structural robustness of interconnected network models. This paper presents a framework for investigating the structural robustness of interconnected network models. In this context, we define functional nodes within interconnected networks as those belonging to clusters of size greater than or equal to $s$ in the local network, while maintaining at least $M$ significant dependency links. This model presents precise analytical expressions for the cascading failure process, the proportion of functional nodes in the stable state, and a methodology for calculating the critical threshold. The findings reveal an abrupt phase transition behavior in the system following the initial failure. Additionally, we observe that the system necessitates higher internal connection densities to avert collapse, especially when more effective support links are required. These results are validated through simulations using both Poisson and power-law network models, which align closely with the theoretical outcomes. The method proposed in this study can assist decision-makers in designing more resilient reality-dependent systems and formulating optimal protection strategies.
 \end{abstract}
\maketitle

\section{1. Introduction}
 \fontsize{12}{16}\selectfont
   In recent decades, the infrastructure of the power system has become intricately interconnected with a wide range of advanced communication, computational, and control technologies. These infrastructure networks, including the power grid and communication networks, rely on one another to fulfill their functions, and provide mutual support, to ensure the reliable supply of electricity for both commercial operations and daily life. Recognizing its pivotal role, scholars and engineers have collaboratively dedicated several decades to improving the stability, reliability, and efficiency of the infrastructure network system. Through the integration of advanced technologies such as sensors, communication, and control, modern electrical infrastructure networks are evolving into a Cyber-Physical system (CPPS) \cite{b1,b2,b3,b4,b5}. While these innovations have enhanced system capabilities, they have also introduced greater complexity and increased susceptibility to potential attacks. Meanwhile, this also presents a greater challenge for the analysis of the system robustness \cite{b6,b7,b8,b9}.

   Due to the interdependence of the communication network and the power network within this infrastructure system, a system failure has the potential to trigger cascading failures on a significant scale, which could ultimately result in catastrophic consequences. For example, in 2015, Ukraine experienced a blackout caused by hackers who used malware to disconnect circuit breakers in the country's power system through its communications network, resulting in thousands of customers losing access to electricity \cite{b10}. Additionally, a blackout struck India in July 2012 when a faulty AC transmission line tripped, triggering a chain reaction of voltage crashes that ultimately led to a massive power outage. More than 620 million people were affected as half of the country experienced a blackout lasting over two days \cite{b11,b12}. These incidents emphasize the critical need to safeguard the security and robustness of power networks, which is essential for both national safety and the well-being of citizens \cite{b13,b14,b15}. This also highlights the significance of understanding cascading failure mechanisms and failure propagation  in multiple support networks \cite{b16,b17,b18,b19,b20,b21}.
    
   Network robustness, which assesses the capacity to maintain essential functions in the face of attacks on network components, serves as a metric for evaluating system performance. Analyzing robustness based on topological structure effectively reflects the influence of network connectivity on robustness. Dong et al. characterized network robustness by studying the percolation which interacting networks with feedback-dependency links \cite{b22}. Buldyrev et al. introduced a theoretical framework model to analyze the robustness of one-to-one dependent networks subjected to cascading failures based on the power network and an Internet network in Italy, from the perspective of complex networks \cite{b23}. Chen et al. integrated game theory with network science to explore the impact of heterogeneous interdependent networks on system robustness \cite{b24}. Dong et al. explored the resilience of network systems through percolation in interdependent networks with feedback-dependency links \cite{b25}. Yuan et al. studied the robustness of interdependent networks through one-to-one interdependence between single nodes \cite{b26}. Additionally, assortativity, disassortativity, and local interdependence across different networks, has a significant impact on the structural robustness of the system \cite{b27,b28,b29}. 

   In real scenarios, communication network base stations necessitate multiple sources of stable and reliable power supply from electrical distribution sites to maintain their normal operation. And power network sites rely on monitoring and information transmission from multiple sites in the communication network for remote control and failure monitoring. When electrical distribution sites initially experience failure, communication base stations lose their power supply, resulting in communication interruptions. This affects the stable supply of electricity and disrupts the normal operation of the entire system \cite{b30,b31,b32,b33,b34,b35}. Furthermore, due to the distributed multiple dependencies between systems, during cascading failures, the functionality of nodes relies not only on the connectivity of the network itself but also on the multiple supports in its coupled system after undergoing the initial failure \cite{b36,b37,b38,b39,b40,b41,b42,b43}. 

   However, the robustness of distributed coupling systems with multiple support in real-world scenarios remains poorly investigated. To address these research gaps, we propose an analytical framework that describes the cascading failure mechanism of the system through both theoretical and numerical simulations. We here introduce finite component size $s$ and effective external support edges $M$ as measures to quantify the influence of each component on network robustness.
   
   This paper mainly focus on studying the effects of both network cluster size (local connectivity) and external network support relationships on the system robustness. Overall, the major contributions of this paper are as follow:
   \begin{itemize}
  \item[$\bullet$] Based on real coupling scenario, this paper proposes a coupled network system with multiple dependent support, and defines the cascading failure mechanism of the system after suffering a failure, from the perspective of complex networks.
  \item[$\bullet$] A theoretical framework is developed to evaluate the system robustness by considering impact of cluster size and the existence of multiple external effective dependency links.
  \item[$\bullet$] Based on the proposed simulation and theoretical framework, we evaluate the impact of different network topologies on increasing robustness.
  \item[$\bullet$] The finding highlight that the cluster size and effective dependency links both affect the robustness of the system, and the result provide practical insights for building robust coupling systems.
\end{itemize}

   The paper is organized to five sections. Section 2 presents the robustness analysis framework of the coupled network. In Section 3, we delve into the theoretical modelling of cascade failure. Section 4 is dedicated to analyze the robustness of power and communication system, taking several critical factors into account through a lot of numerical simulation. Finally, we present our conclusions in Section 5.

   \section{2. Robustness analysis framework of the coupling network}
\begin{figure}[ht]
   \begin{center}
        \includegraphics[width=6.3in]{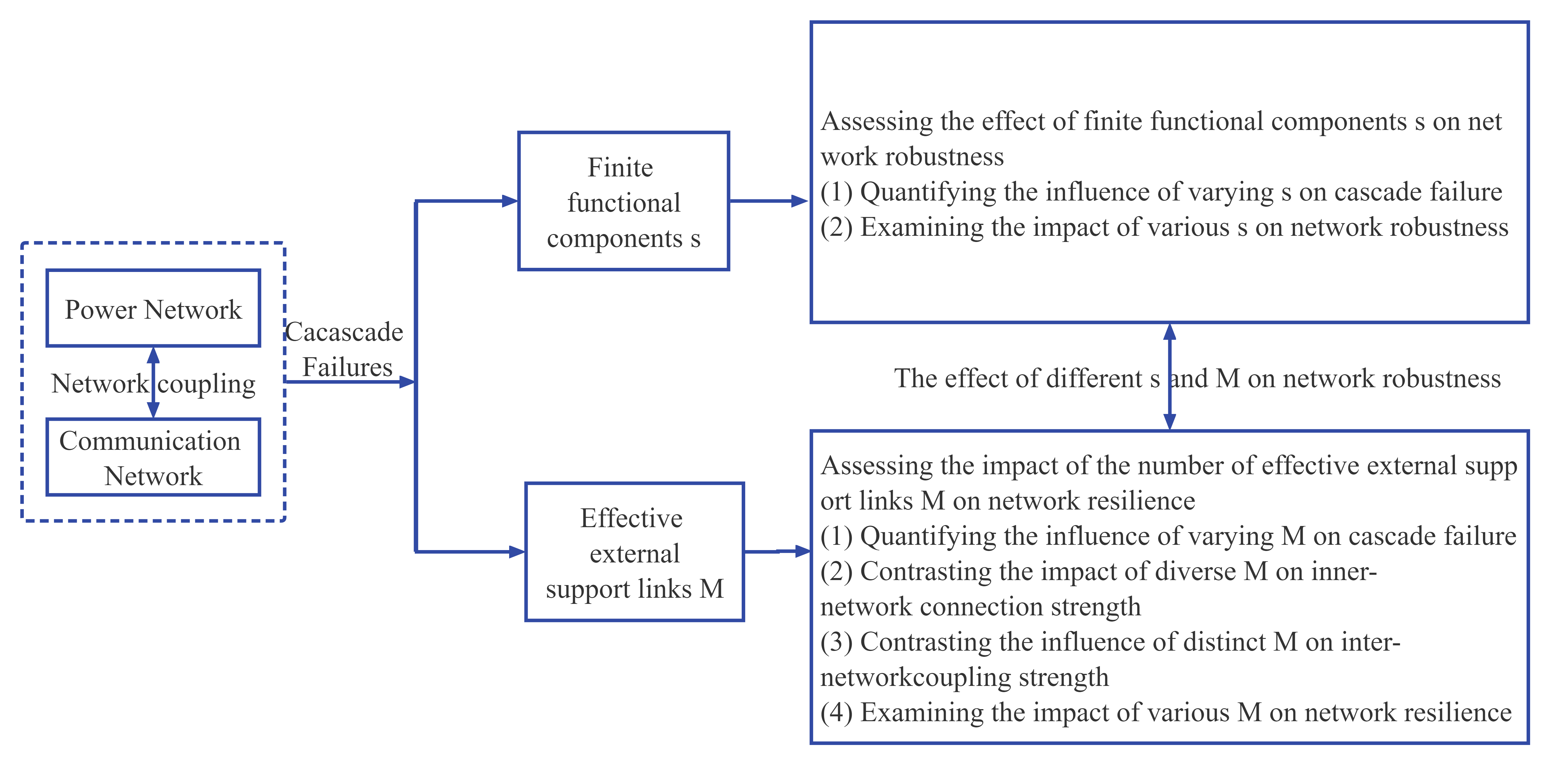}
        \hspace{2mm}\vspace{0mm}
    \end{center}
    \begin{spacing}{1.5}
    \captionsetup{font=small}
    \caption{
    \fontsize{12}{16}\selectfont
    A framework for  the structural robustness of coupling networks }
    \label{fig1}
    \end{spacing}
\end{figure}

    Our framework for assessing the robustness of coupling network structures, utilizing real-world scenarios from Cyber-Physical system, is depicted in Fig.~\ref{fig1}. We first construct coupling networks with Cyber-Physical system as the actual background. Then we simulate cascade failure failures by applying random attack strategies to the constructed coupled networks. Finally, we consider the effect of effectively supported connecting edges between networks and the size of components within a network on the robustness of the network.

    To fill the gaps within the pre-existing model frameworks, we introduce two sub-networks denominated as $A$ and $B$, characterized by degree distributions $P_A(k)$ and $P_B(k)$, respectively. These two subnets exhibit mutually interdependent relationship, wherein the support furnished by $A$ to $B$ conforms to the distribution $\widetilde{P}_A(\widetilde{k}_A) $, and conversely, the support provided by $B$ to $A$ is expressed as $\widetilde{P}_B(\widetilde{k}_B)$.

     \begin{figure}[ht]
   \begin{center}
        \includegraphics[width=6.3in]{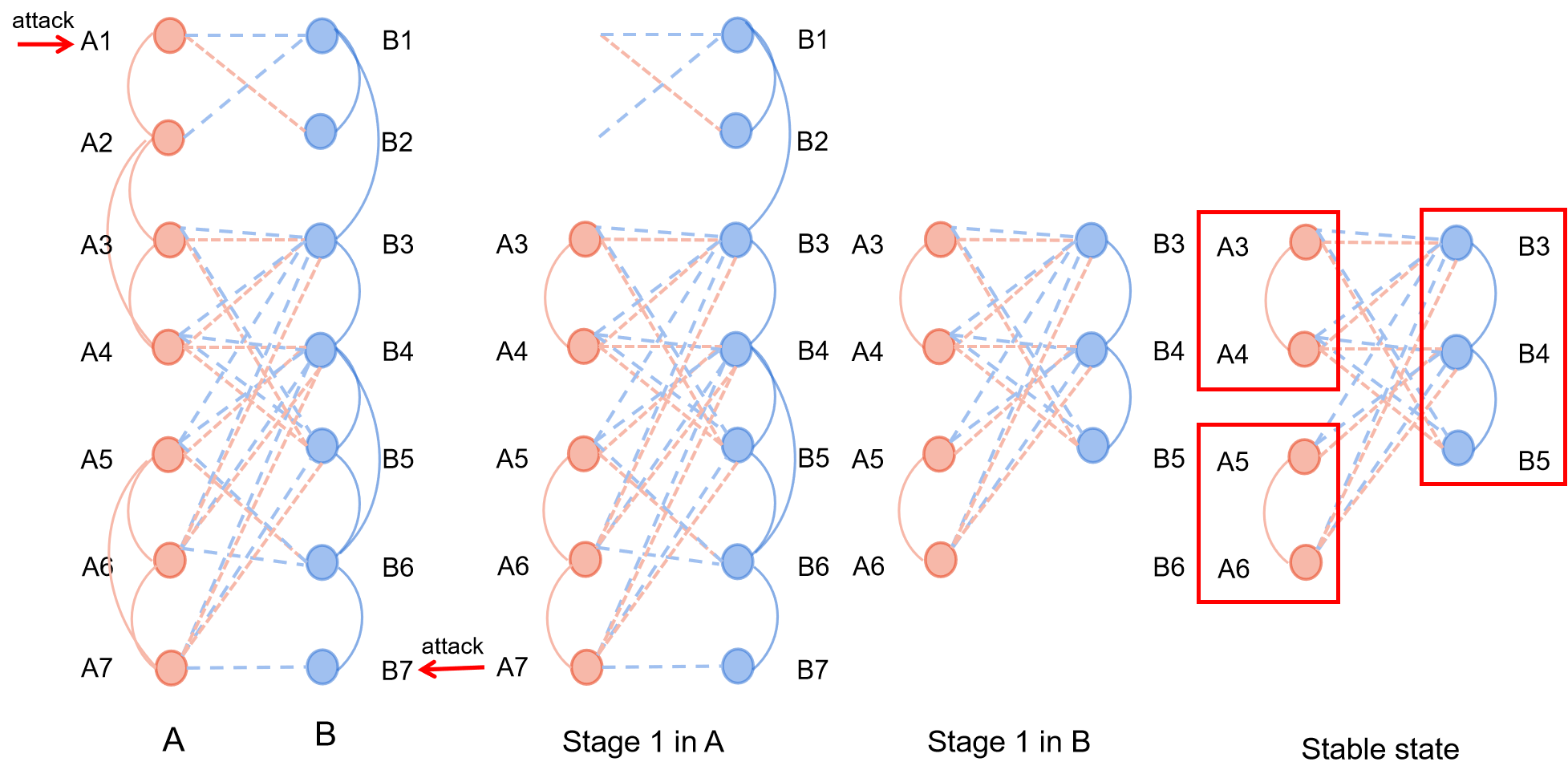}
        \hspace{2mm}\vspace{0mm}
    \end{center}
    \begin{spacing}{1.5}
    \captionsetup{font=small}
    \caption{
    \fontsize{12}{16}\selectfont
    Schematic illustration of cascading failure within a coupled network characterized by multiple dependent links, studied from the finite components perspective and taking the network with $N_A=N_B=7$ as an example.The curve and dash lines describe connectivity links within both networks and support links between networks respectively. (a) At the initial stage, nodes $A1$ and $B7$ are initially attacked (red arrows) and become nonfunctional. (b) Since node A2 does not belong to an association of size $s>=2$, it also fails. (c) The failure of these nodes causes the number of valid supporting edges of node $B1$,$B2$,$B6$ to be less than $M=2$ in network $B$, and thus these three points fail. This process is repeated until all nodes in the network satisfy the two conditions for a valid node, eventually reaching the steady state (d).}
    \label{fig2}
    \end{spacing}
\end{figure}
  Due to the multilateral dependency of the nodes, for example, a communication station must be supported by multiple power stations to ensure stable operation. It is assumed that each effective node must have at least $M$ effective support edges here. At the same time, due to the development of science and technology, the actual network scale expansion, the finite connectivity group can not be ignored, here will be coupled network in the effective nodes belong to the largest connectivity group of the condition is relaxed to its belong to the effective connectivity group, i.e., the size of the associations larger than $s$. Therefore, the condition for a node to be effective in this type of dependent network is: (1) the functional node belongs to a cluster within the network with a size no smaller than; (2) It has at least $M$ effective support links connected to functional nodes of other networks.

\section{3. Theoretical Modelling of cascade failure}
 \fontsize{12}{16}\selectfont
   
   When system components fail, they are promptly removed from the system. The removal of these failed components not only changes the topology of the coupling system but also impacts its function. For instance, the operational status of all components will change due to alterations in the system's connectivity, potentially leading to disruptions in the sensing, computing, and control processes of the dispatching. During the cascading failure phase, the assessment focuses initially on network $A$, and subsequently on network $B$, without the latter influencing the final result. As the network's nodes undergo processing in the $n$-th stage within network $A$, we posit that all supporting nodes within network $B$, which proved effective during the preceding ($n-1$) stage, remain available, and vice versa. Within this framework, the portion of presently operational nodes in networks $A$ and $B$ is represented as $S^A_n$ and $S^B_n$, respectively.
   
\subsection{3.1 External support links}
   We assume that component within the respective  network $A$ and $B$ sustain damage, resulting in the random removal of a fraction denoted as $(1-p^A)$ and $(1-p^B)$ from networks $A$ and $B$.  After failure a functional node not only needs to have at least $M$ of support links from another network, but also belong to a cluster of size greater than or equal to $s$ in the local network. During this process, the fraction of effective support links, each comprising at least $M$ links, within networks $B$ and $A$ during the $n$th stage are designated as $y^A$ and $y^B$, respectively. These values are computed using the following formulas:
   \begin{equation}
   \left\{
   \begin{aligned}
       &y^A=p^A(1-r^A_{n,M}),
       \\
       &y^B=p^B(1-r^B_{n,M}).
    \end{aligned}
    \right.
    \label{eq1}
    \end{equation}
    Where $1-r^A_{n,M}$ and $1-r^B_{n,M}$ denote the probabilities of possessing at least $M$ effective support links from networks $B$ and $A$ during the $n$th stage, in networks $A$ and $B$, correspondingly. To further elucidate, we use $S^A_{n,M,s}$ and $S^B_{n,M,s}$ to denote the proportion of effective nodes in the $A$ and $B$ networks at the $n$th stage, respectively \cite{b22}.

\begin{equation}
    \begin{aligned}
    1-r^A_{n,M}=1&-r^A_{n,M-1}\\
    &-\sum_{\widetilde{k}_A=M-1}^{\infty}{\left(\begin{matrix}\widetilde{k}_A\\M\\\end{matrix}\right)\widetilde{P}_A(\widetilde{k}_A-M+1)(1-S^B_{n-1,M,s})^{\widetilde{k}_A}}(S^B_{n-1,M,s})^{(M-1)}.
    \end{aligned}
    \label{eq2}
\end{equation}
  $1-r^A_{n,M}$ denotes the probability that a node in the nth stage has at least $M$ valid supporting links in the network $B$ ($A$). The last term is expressed as the fraction of nodes in network $A$ ($B$) that have only $M-1$ effective support links. 

   The vulnerability profile of the coupling network following an attack, without factoring in the local network, can be derived using Eq.~\ref{eq2}. It becomes evident that as the parameter $M$ increases, the coupling network becomes increasingly susceptible to collapsing under attack, resulting in compromised efficiency and suboptimal performance.

\subsection{3.2 Modelling of cluster size}

   In practice, the normal operation of the coupling system requires not only the support of the inter-network, but also the support of the system itself. Next, we discuss the effect of the size of the finite components on the robustness of the system.
   We define $g^A_{n,M}(y^A)$ and $g^B_{n,M}(y^B)$ as the fractions of nodes within the giant component in relation to the surviving nodes during the $n$th stage in networks $A$ and $B$, respectively \cite{b41}.
\begin{equation}
\left\{
\begin{aligned}
    &g^A_n(y^A)=1-G_A[1-y^A(1-f_A)]
    \\
    &g^B_n(y^B)=1-G_B[1-y^B(1-f_B)]
    \end{aligned}
    \right.
    \label{eq3}
\end{equation}
   Where $g^A_{n,M}(y^A)$ and $g^B_{n,M}(y^B)$ as the fractions of nodes within the giant component in relation to the surviving nodes during the $n$th stage in networks $A$ and $B$, respectively. $G_A(x) = \sum_{k=0}^{\infty}{P_A(k)x^k}$ and $G_B(x) = \sum_{k=0}^{\infty}{P_B(k)x^k}$ are the generating functions of networks $A$ and $B$ with degree distributions $P_A(k)$ and $P_B(k)$, respectively, and $f_A(f_B)$ is the probability that a node is not in the giant component. It satisfies a recursive equation \cite{b35}
\begin{equation}
\left\{
\begin{aligned}
    &f_A=H_A[1-y^A(1-f_A)],
    \\
    &f_B=H_B[1-y^B(1-f_B)],
    \end{aligned}
    \right.
    \label{eq4}
\end{equation}

   where $H_A(x)=\frac{G'_A(x)}{G'_A(1)}$ ($H_B(x)=\frac{G'_B(x)}{G'_B(1)}$). When the size of the giant component(1-$f_A(f_B)$) approaches zero, the Cyber-Physical system itself is on the verge of collapse. We also calculate the generating function for the component size distributions \cite{b42}.
\begin{equation}
C(x,y)=\sum_{s=1}^{\infty}{\pi_s(y)x^s}=xG[B(x,y)y+1-y],
\label{eq5}
\end{equation}

    where $\pi_s(y)$ is the fraction of nodes in a component of size $s$ in network $A(B)$ relative to the surviving nodes with fraction $y$, and $B(x,y)$ satisfies the recursive equation \cite{b42}
\begin{equation}
    B(x,y)=xH[B(x,y)y+1-y].
    \label{eq6}
\end{equation}
   It is worth noting that when $x=1$, Eq.~\ref{eq7} is equivalent to Eq.~\ref{eq5}, and hence,
\begin{equation}
    C(1,y)=\sum_{s=1}^{\infty}{\pi_s(y)}=1-g(y).
    \label{eq7}
\end{equation}

 Using the Lagrange inversion formula, we obtain the coefficients $\pi_s(y)$ when $s>1$ \cite{b42}:
\begin{equation}
    \pi_s(y)=\frac{y\langle k \rangle}{(s-1)!}\frac{d^{s-2}}{dx^{s-2}}[H(xy+1-y)]^s|_{x=0},
    \label{eq8}
\end{equation}
and
\begin{equation}
    \pi_1(y)=G(1-y).
    \label{eq9}
\end{equation}

     Instead of the function $g^i(y^i)$, we use the function $g^i_{n,s}(y^i)$, which is defined in the same way as $g^i(y^i)$, only the giant component is replaced by a component of size greater than or equal to $s$
\begin{equation}
\left\{
\begin{aligned}
    &g^A_{n,s}(y^A)=1-\sum_{r=1}^{s-1}{\pi_{A,s}(y^A)},
    \\
    &g^B_{n,s}(y^B)=1-\sum_{r=1}^{s-1}{\pi_{B,s}(y^B)}.
    \end{aligned}
    \right.
    \label{eq10}
\end{equation}

    Notably, it becomes evident that with the increase in cluster size $s$, a higher network critical threshold point emerges in the event of an attack on the coupling network. This heightened threshold indicates an increased likelihood of triggering complete network failure, thereby contributing to compromised efficiency and suboptimal performance.

\subsection{3.3 Effective nodes in the cascade process}
   In our model, we arrest the cascade failure process through simulation of the coupling network under attack, contingent upon the functional nodes within the coupling network fulfilling the subsequent criteria: (i) being part of a cluster $s$ within the coupling network; and (ii) possessing a minimum of $M$ valid support edges in the coupling network. Throughout this process, the count of presently operational nodes diminishes over time, ultimately converging to the quantity of functional nodes that meet both conditions (i) and (ii).

   Hence, we use the generating function approach to derive analytic formulas for $S^A_{n,M,s}$ and $S^B_{n,M,s}$:
   \begin{equation}
    \begin{cases}
     S^A_{M,s}=y^Ag^A_{n,s}(y^A),
     \\
     S^B_{M,s}=y^Bg^B_{n,s}(y^B).
    \end{cases}
    \label{eq11}
    \end{equation}

   The critical thresholds correspond to marginal transition points wherein the networks transition from having finite values to zero, signifying the transformation of the coupling network from a state of normal operation to complete failure. These pivotal junctures can be determined by identifying the intersections of the two curves, $S^A_{\infty,M,s}$ and $S^B_{\infty,M,s}$, through the utilization of Eq.~\ref{eq14}. The critical threshold can be calculated as follows:
\begin{equation}
\frac{d{S^A_{\infty,M,s}}}{d{S^B_{\infty,M,s}}}\times\frac{d{S^B_{\infty,M,s}}}{d{S^A_{\infty,M,s}}}=1.
\label{eq12}
\end{equation}

\section{4. Simulation results and discussion}
\fontsize{12}{16}\selectfont

   This part will analyse the effects of network topology configuration parameters, effective node conditions, and so on, to the system robustness through two common network models, ER and SF, which are used to further verify the reasonableness of the model framework. Without loss of generality, it is set that the two networks have the same size $N_A=N_B=N$, and the average within the two networks and the average degree of the support network between the two networks respectively keep the same $<k_A>=<k_B>=<k>$, $k^A_{inter}=k^B_{inter}=k_{inter}$. Also, each set of results was simulated independently more than 100 times to reduce the errors due to randoms.  
   
\subsection{4.1 Robustness of ER coupled network}

   \begin{figure}[ht]
    \centering
   
        \includegraphics[width=3.3in]{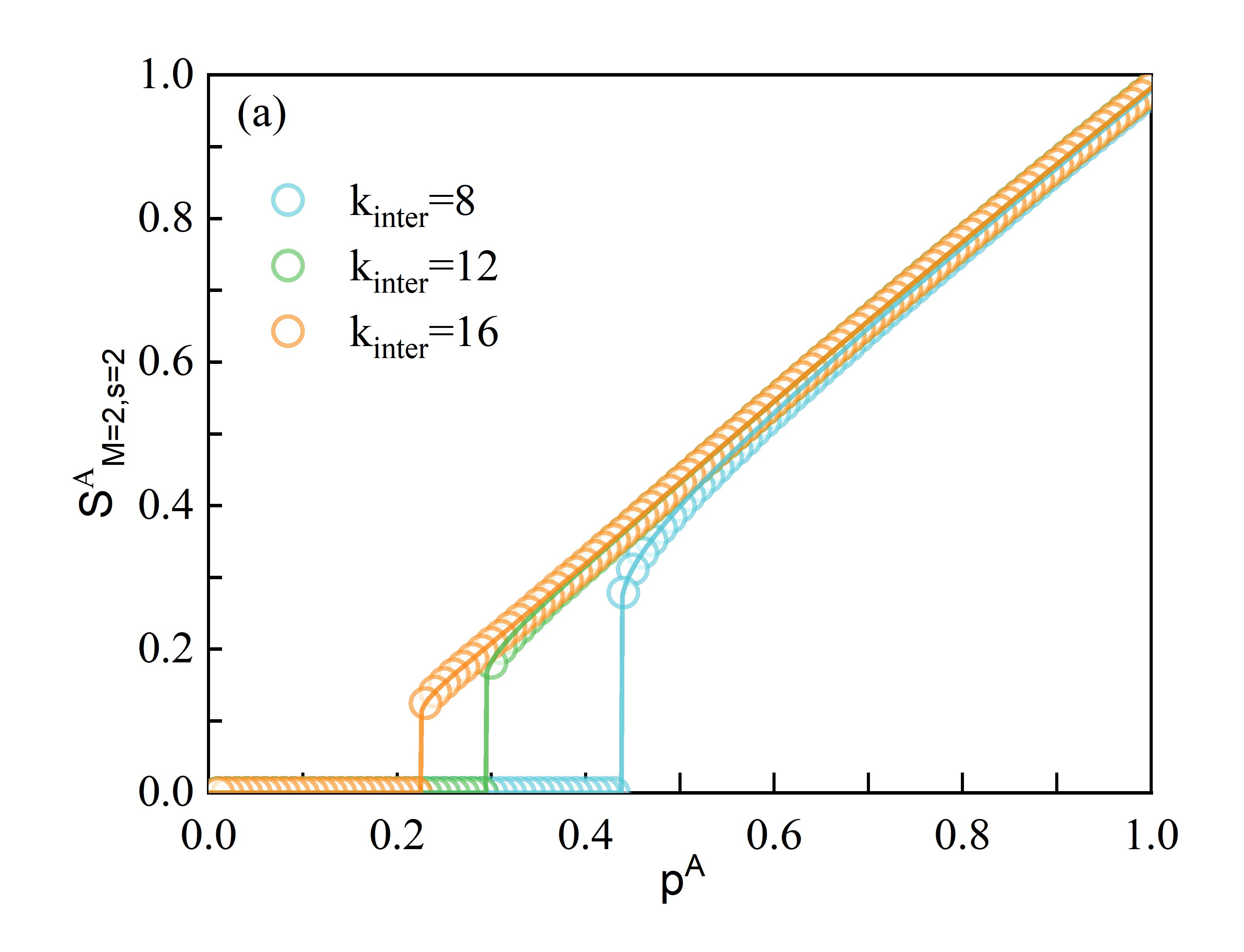} 
        \includegraphics[width=3.3in]{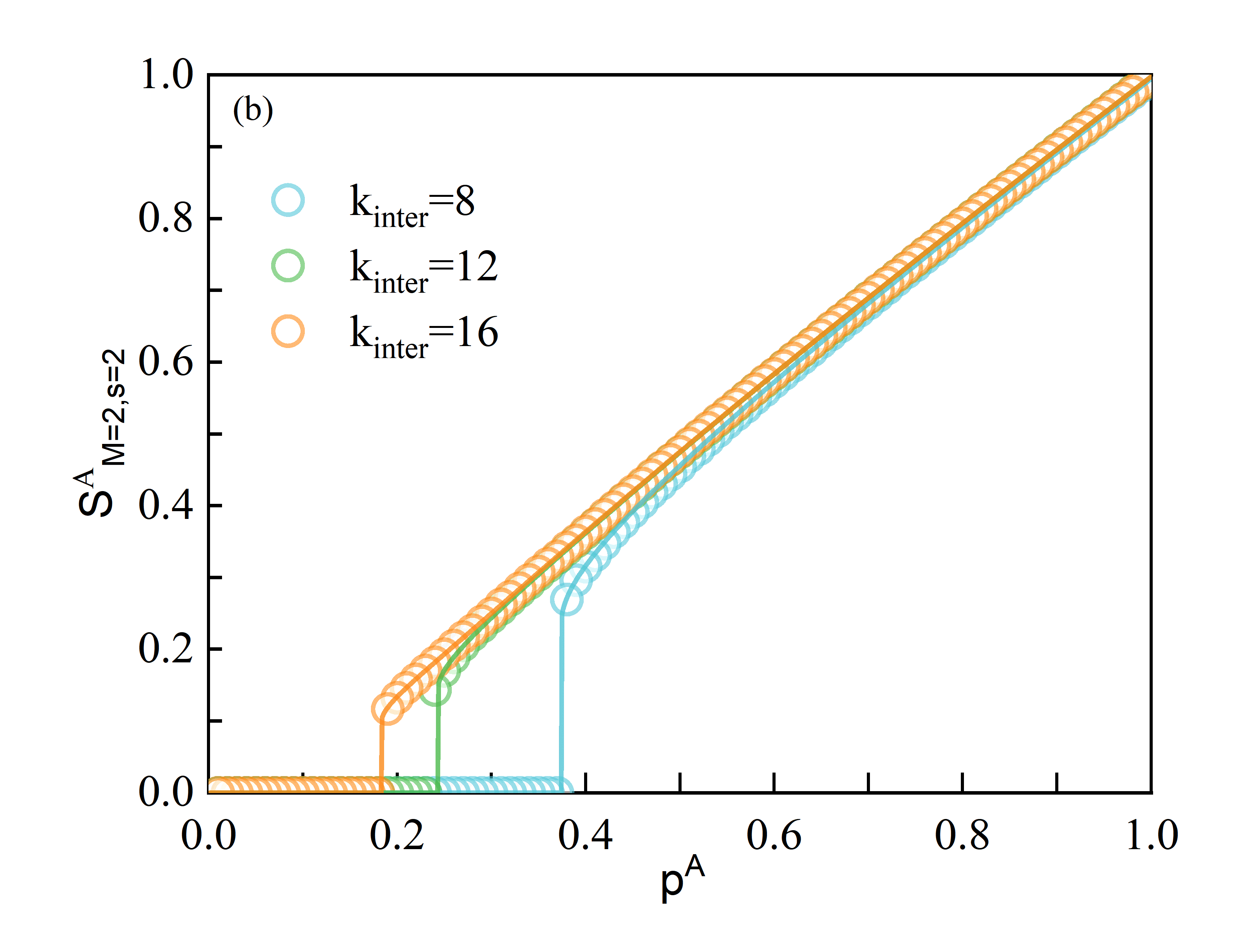} 
        \includegraphics[width=3.3in]{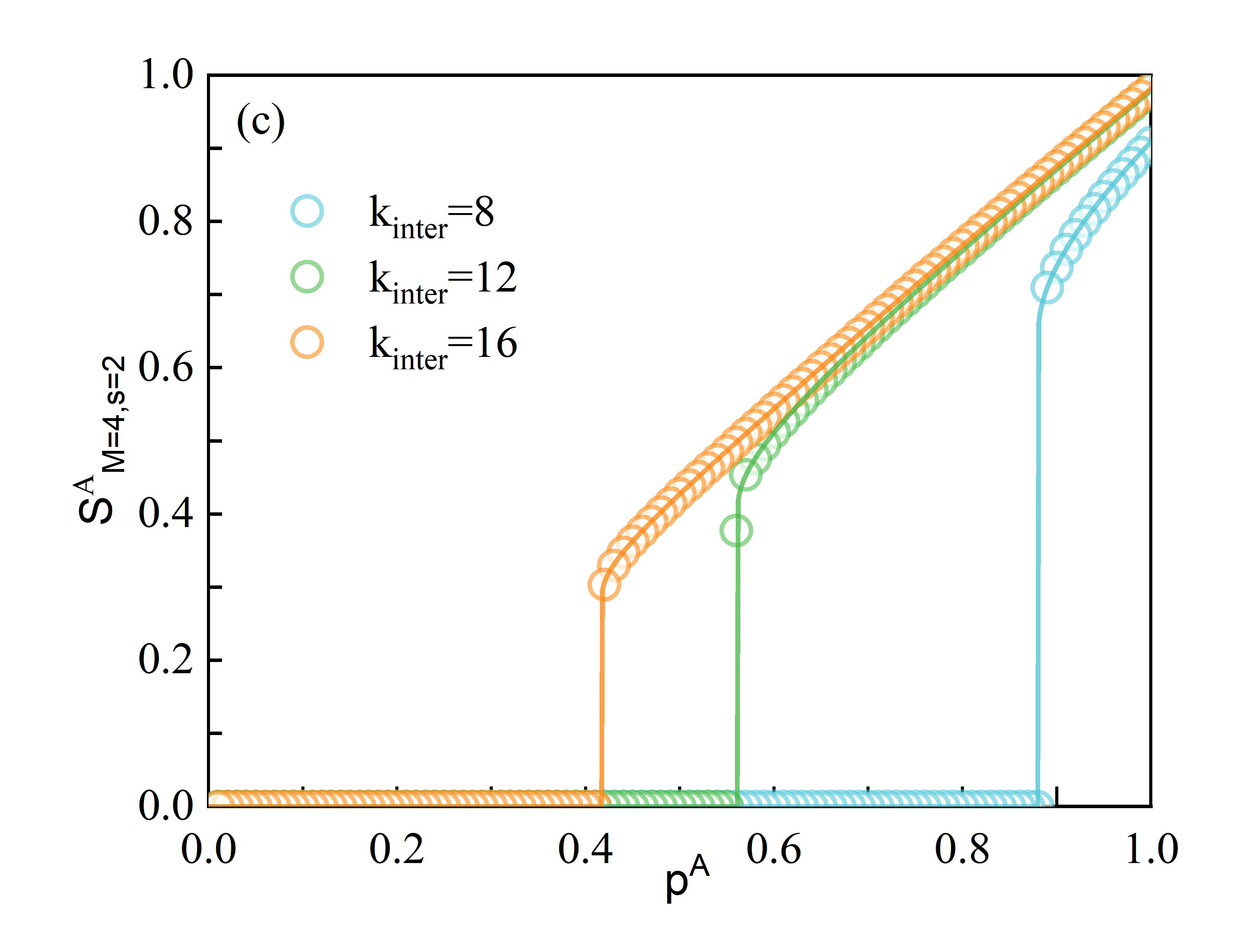} 
        \includegraphics[width=3.3in]{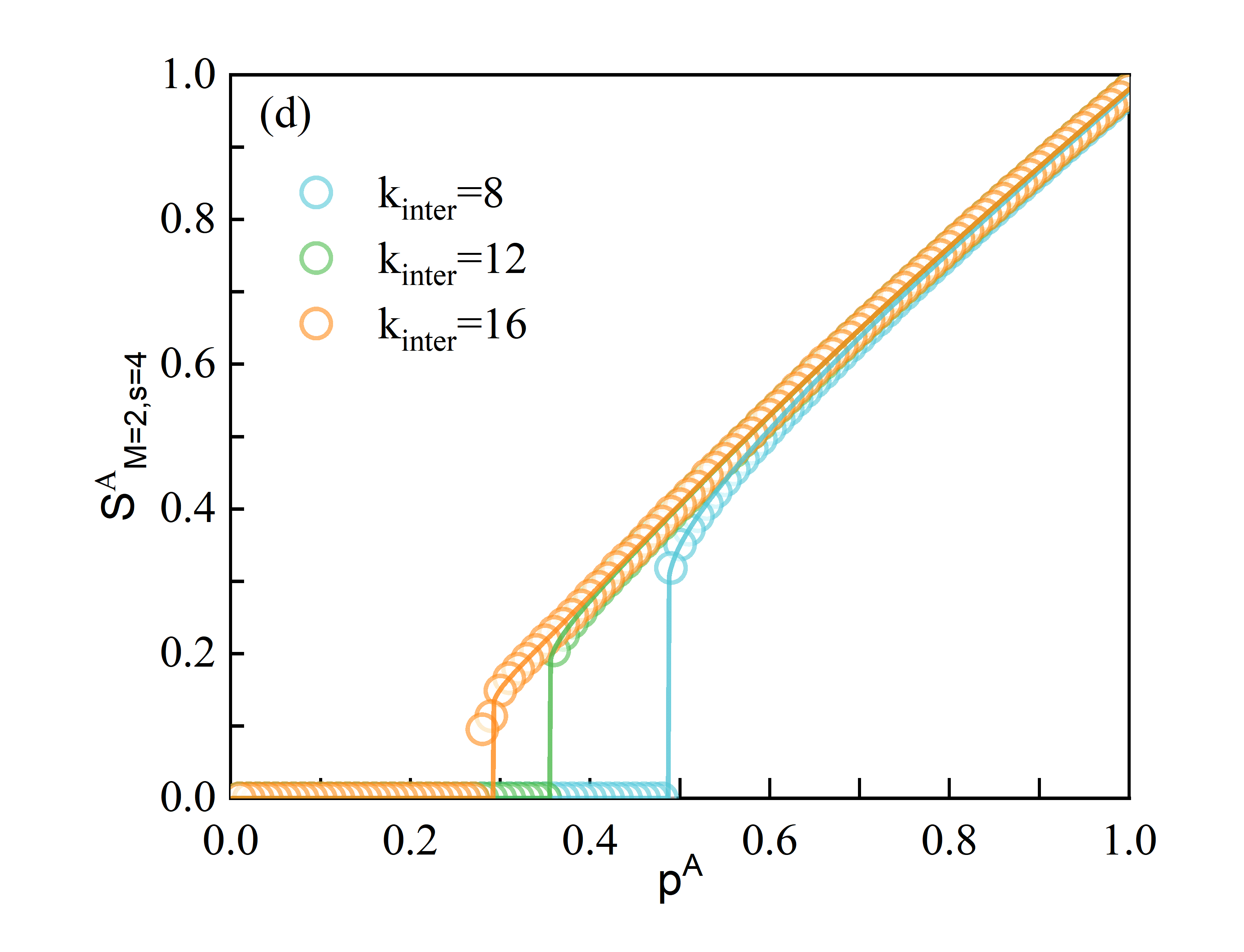}
    
   \begin{spacing}{1.5}
    \captionsetup{font=small}
    \caption{
    \fontsize{12}{16}\selectfont
     $S^A_{M,s}$ as a function of $p^A$ for different parameters of $M$, $s$, and $K_{inter}$: (a) $M=2$, $s=2$,$<k>=4$; (b) $M=2$, $s=2$,$<k>=6$; (c) $M=4$, $s=2$, $<k>=4$; (d) $M=4$, $s=4$, $<k>=4$. Here, the network size is $N_A=N_B=10^6$, average degree of each sub-network is $<k_A>=<k_B>=<k>$,  $k^A_{inter}=k^B_{inter}=k_{inter}$ and $1-p^A=2(1-p^B)$. Circles represent simulation results and solid lines represent theoretical analysis results. }
    \label{fig3}
    \end{spacing}
\end{figure}
    The ER network adheres to a Poisson degree distribution, and its generating function can be expressed as $G(x)=H(x)=e^{\langle k \rangle (x-1)}$, so Eq.~\ref{eq8} can be rewritten as:
\begin{equation}
    \pi_{ER,s}(y)=\frac{(s y \langle k \rangle)^{s-1}e^{-s y \langle k \rangle}}{s!}.
    \label{eq13}
\end{equation}

  By substituting this into Eq.~\ref{eq11}, we can derive:
\begin{equation}
\begin{cases}
S^A_{\infty,M,s}=y^A(1-\sum_{r=1}^{s-1}{\frac{(s y^A \langle k \rangle)^{s-1}e^{-s y^A \langle k \rangle}}{s!}}),
\\
S^B_{\infty,M,s}=y^B(1-\sum_{r=1}^{s-1}{\frac{(s y^B \langle k \rangle)^{s-1}e^{-s y^B\langle k \rangle}}{s!}}).
\end{cases}
\label{eq14}
\end{equation}

   From Fig.~\ref{fig3}(a), it can be observed that, in the ER-coupled network model, the analytical results (lines) are in complete agreement with the simulations (symbols). Furthermore, as the fraction of remaining nodes, $p^A$, increases, Network $A$ exhibits a first-order phase transition in its largest connected component. Fig.~\ref{fig3}(b) demonstrates that, with the increase in the average degree $<k>$, the critical threshold point $p_c^A$ decreases. Additionally, the critical threshold point $p_c^A$ gradually decreases with an increase in the coupling network's average degree $k_inter$. It is evident that in this network model, increasing the support edges effectively enhances the system's robustness. By increasing the number of support edges under the condition of effective nodes, it can be observed that the critical point of Network $A$ also increases, as shown in Fig.~\ref{fig3}(c). As expected, an increase in the number of support edges results in more stringent conditions for effective nodes, subsequently weakening the system's robustness. Fig.~\ref{fig3}(d) illustrates the influence of the size $s$ of the effective connected component on the network. A smaller value of $s$ implies that there are more nodes in communities with sizes greater than or equal to $s$. Hence, the critical point $p_c^A$ increases with an increase in the value of $s$.

   \begin{figure}[ht]
    \centering

        \includegraphics[width=3.3in]{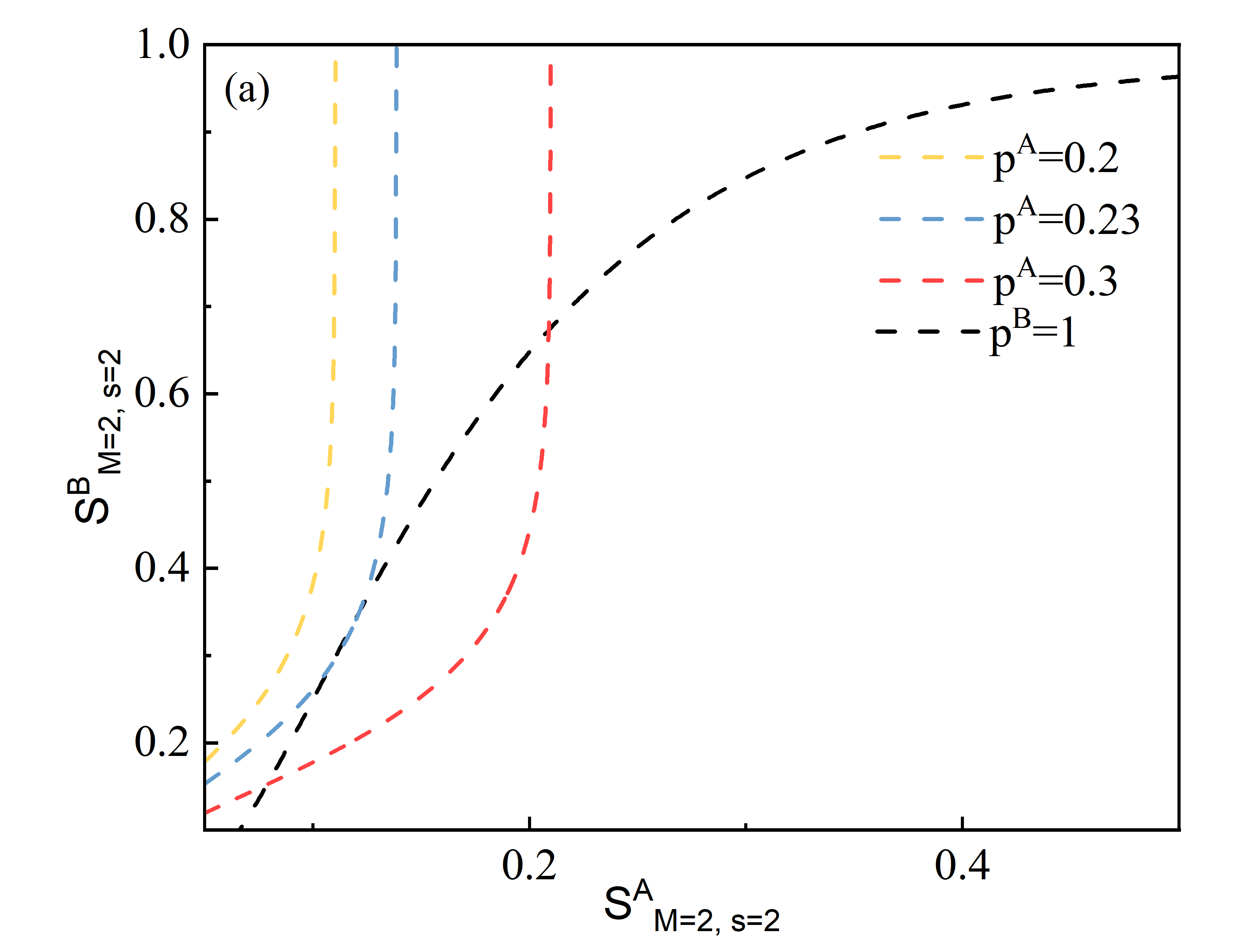}
        \includegraphics[width=3.3in]{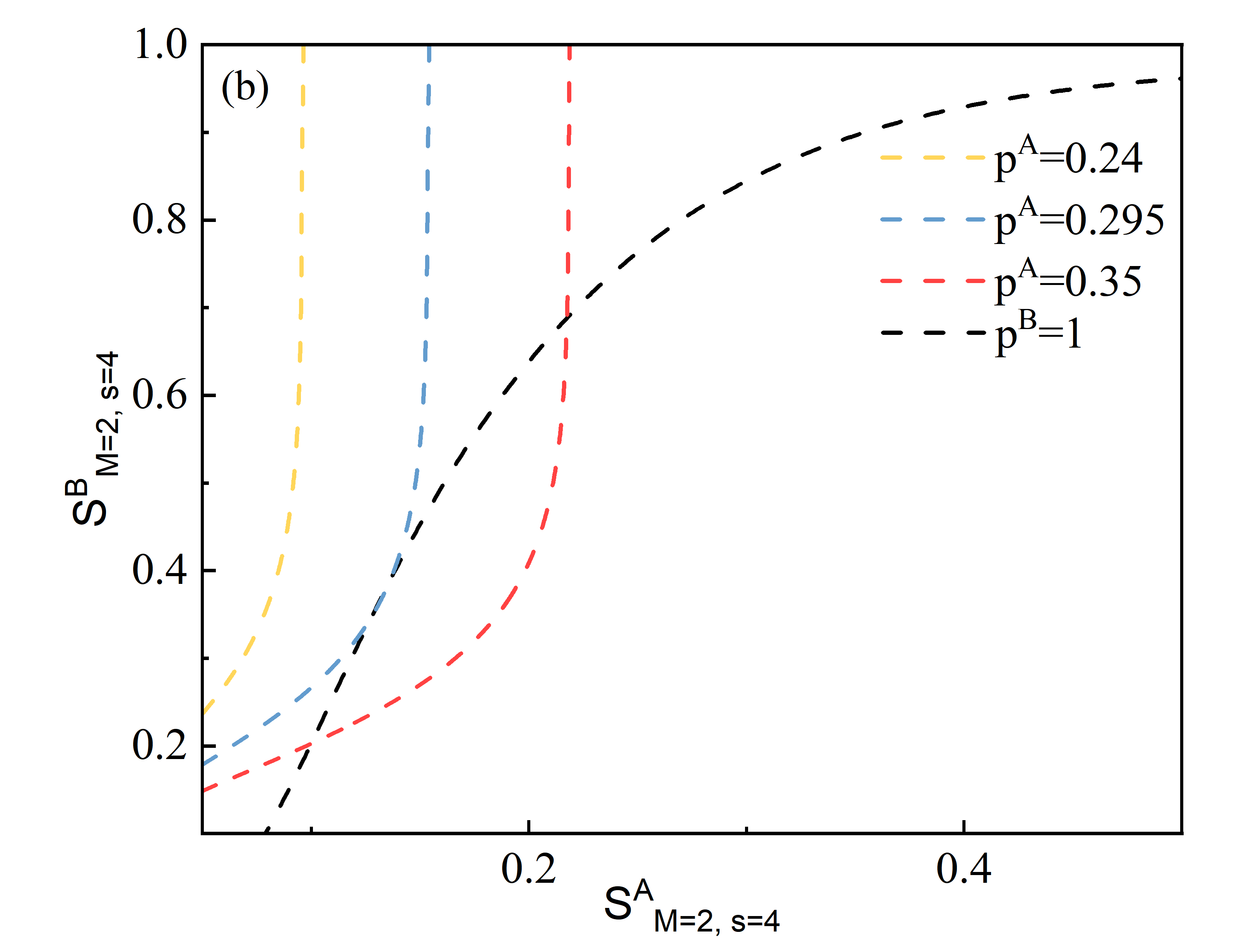} 
    \hspace{2mm}\vspace{0mm}
     \begin{spacing}{1.5}
    \captionsetup{font=small}
    \caption{
    \fontsize{12}{16}\selectfont
    The determination of the critical threshold $p^A_c$ for dependent networks. The determination of the critical threshold $p^A_c$ for couple networks. For the couple ER networks with $\Bar{k}=4$ and $k_{inter}=12$ and $p^B=1$, the relationship between $S^A_{\infty,M,s}$ and $S^B_{\infty,M,s}$ in Eq.~\ref{eq14} was analyzed for different values of $p^A$ and $s$, where (a) $M=2, s=2$ and (b) $M=2, s=4$.}
    \label{fig4}
    \end{spacing}
\end{figure}

   Interestingly, it becomes evident that both $S^A_{\infty,M,s}$ and $S^B_{\infty,M,s}$ exhibit a single non-zero solution present solely at the tangent point across various cluster sizes $s$. This pattern suggests that a giant component manifests at the critical threshold $p^A_c$, as depicted in Fig.~\ref{fig4}. In cases where $p^A > p^A_c$, a non-zero giant component persists within the coupling network. To ascertain an effective value, we choose the larger intersection point, as it inherently surpasses the solution at $p^A_c$. Conversely, when $p^A < p^A_c$, equation solutions exhibit a lack of intersection points. This absence signifies that the coupling network remains in a collapsed state subsequent to cascading failures, specifically when the proportion of surviving nodes within the network falls below $p^A_c$. Moreover, it is noteworthy that across distinct $M$ and $s$ values, the curves $S^A_{\infty,M,s}$ and $S^B_{\infty,M,s}$ each exhibit a sole non-zero solution. It is worth noting that a change in $M$ and $s$ causes a consequent change in the critical threshold $p_c$.

  \begin{figure}[ht]
    \centering
    
        \includegraphics[width=3.2in]{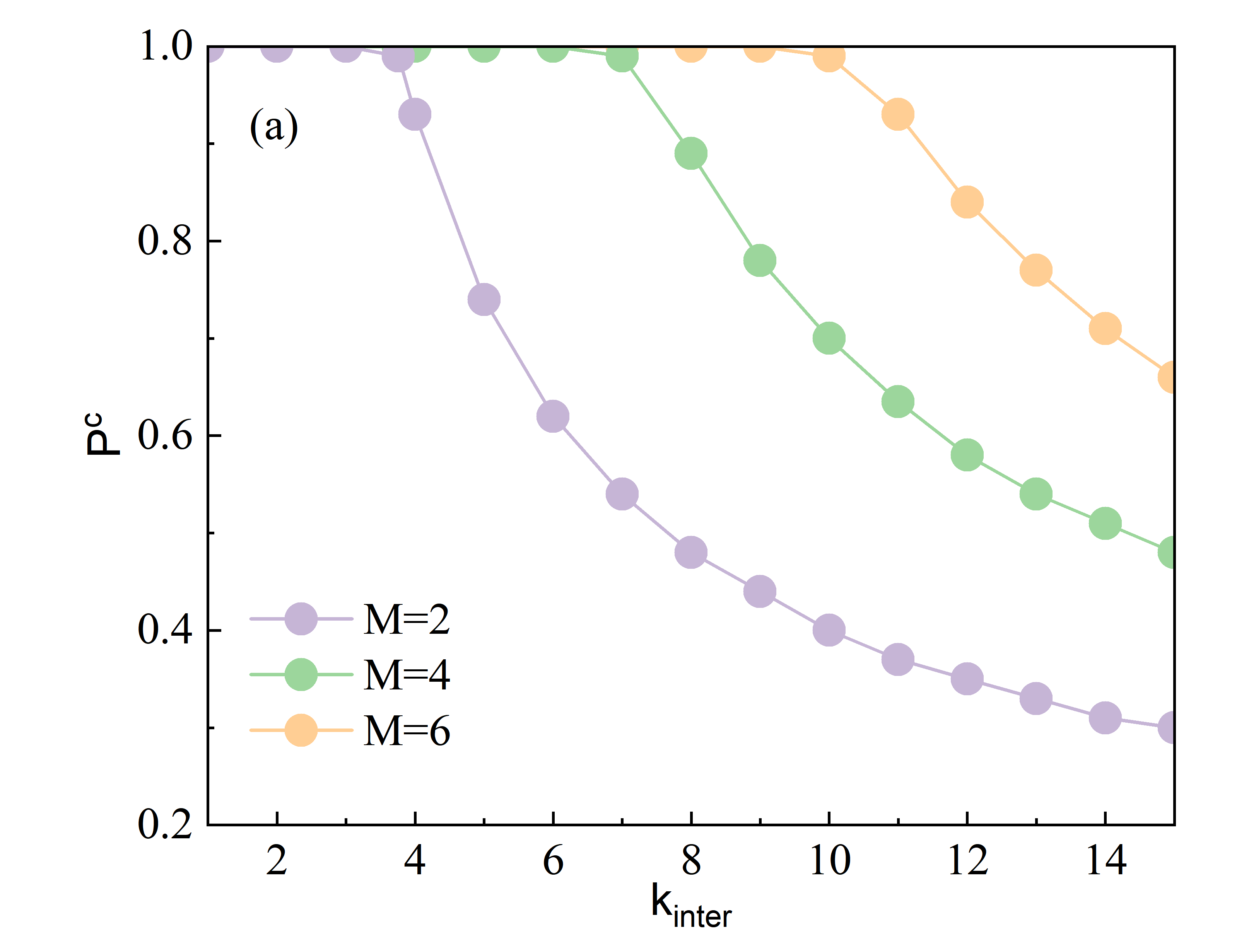}
        \includegraphics[width=3.6in]{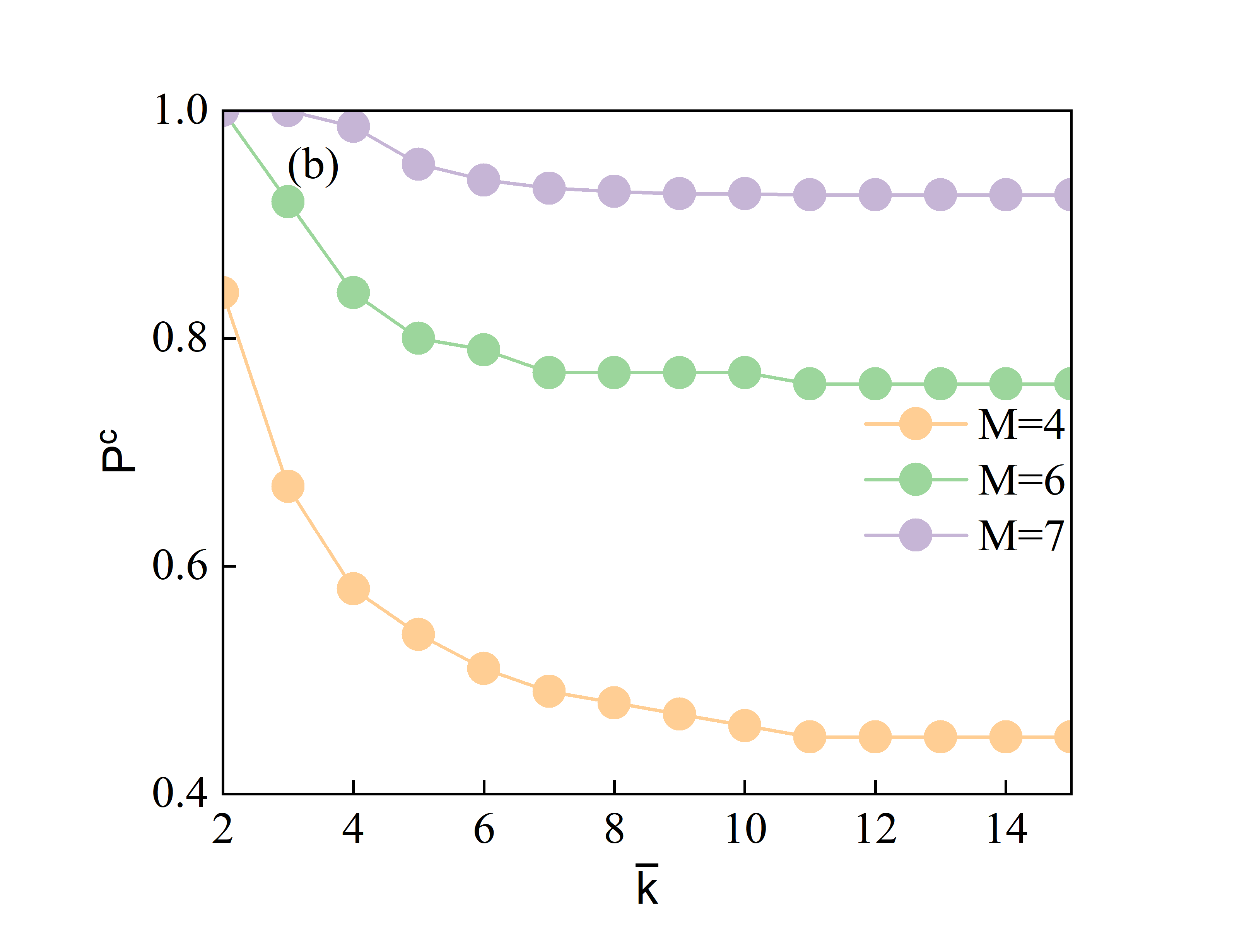}
        \includegraphics[width=3.1in]{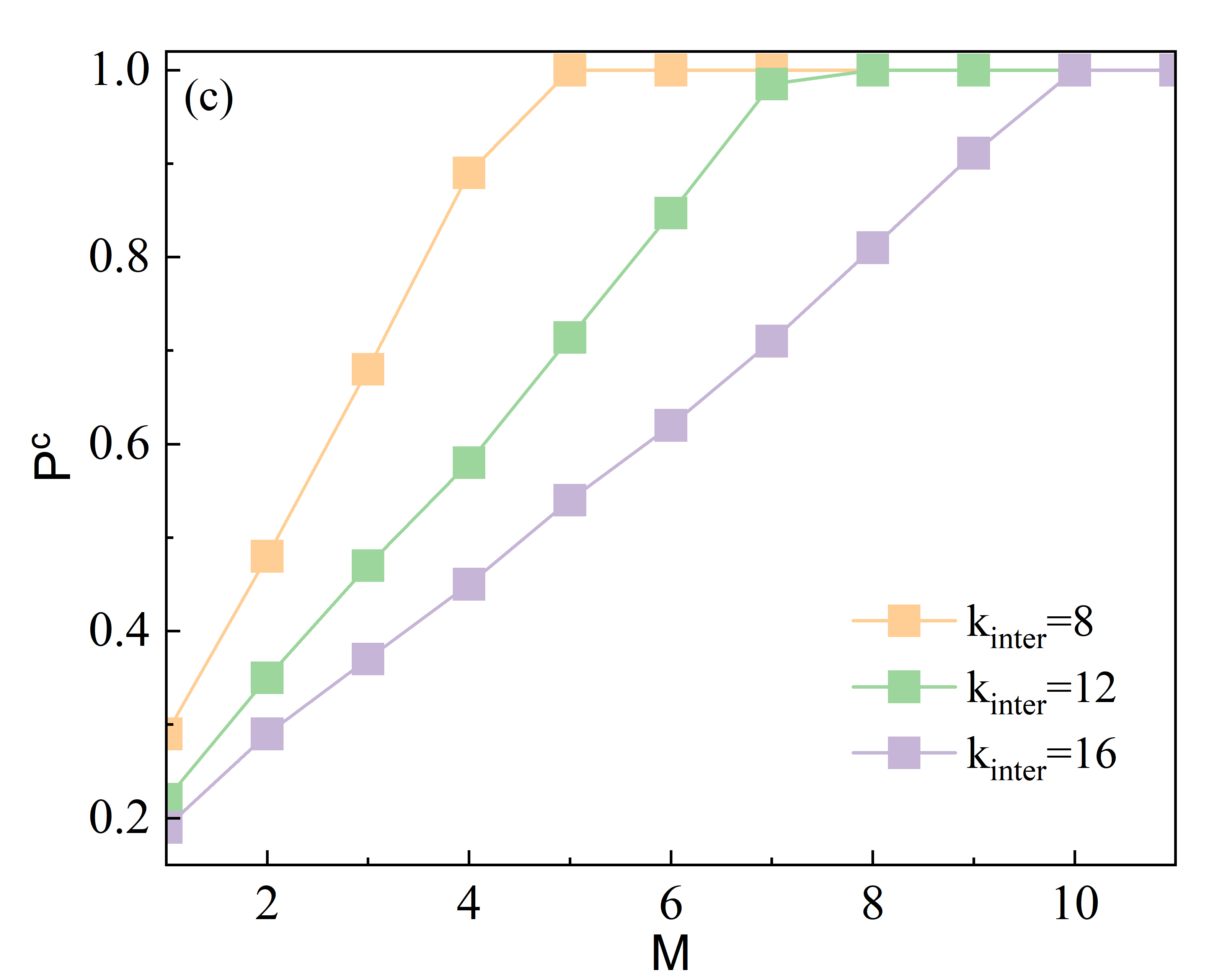}
        \includegraphics[width=3.6in]{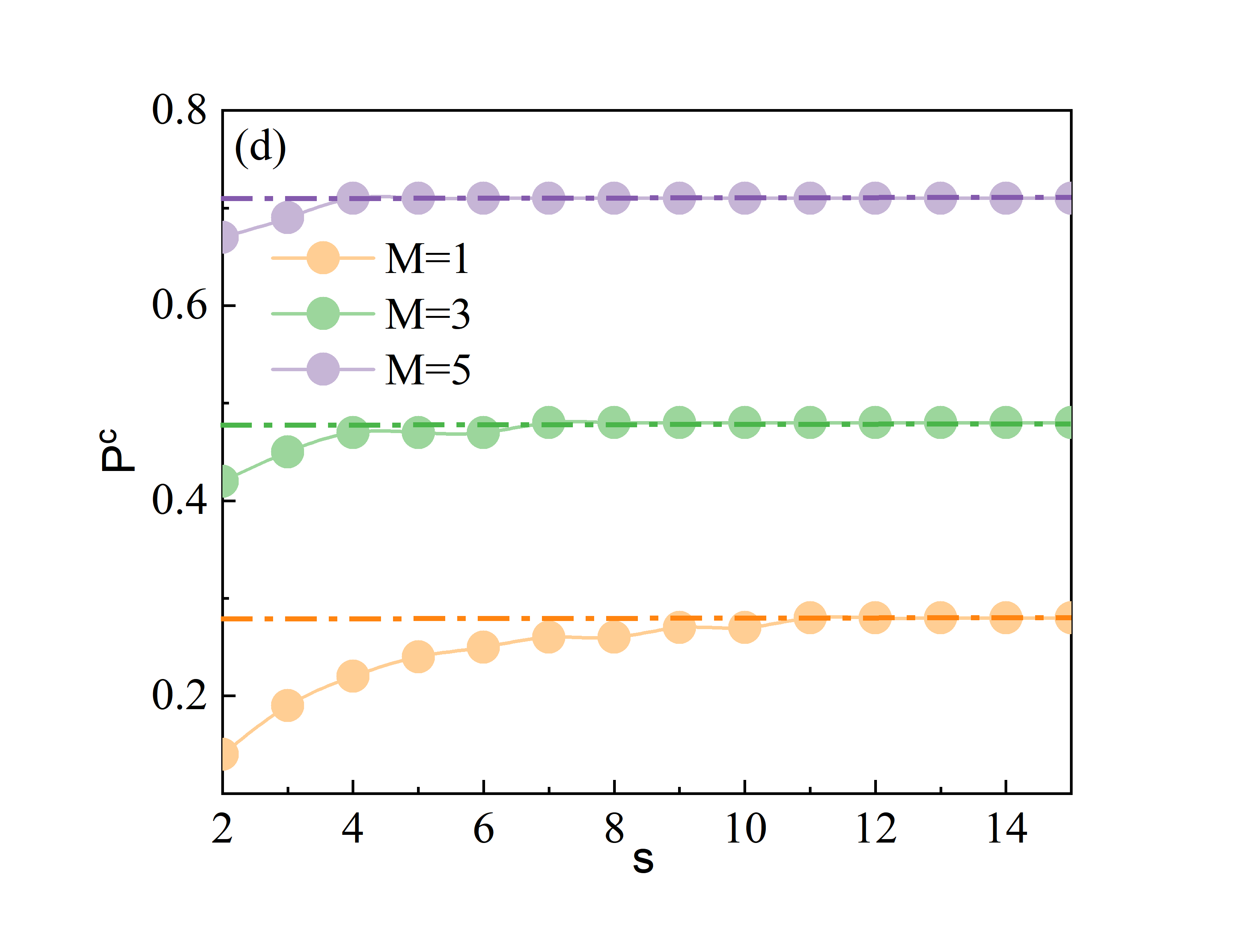}
    \hspace{2mm}\vspace{0mm}
    \begin{spacing}{1.5}
    \captionsetup{font=small}
    \caption{
    \fontsize{12}{16}\selectfont
    Critical points ${p_A}^c$ as a function of different parameters. $1-p^A=2(1-p^B)$. (a) $s=4$ and $\bar{k}= 4$. (b) $s=4$ and $k_{inter}= 12$. (c) $s=4$ and $\bar{k}= 4$. (d) $\bar{k}= 4$, $k_{inter}= 12$.}
    \label{fig5}
    \end{spacing}
\end{figure}

     Eq.~\ref{eq14} and Fig.~\ref{fig3} both indicate that the primary factors affecting this coupled network are the average degree $<k>$ of the sub-network, the average degree $<k_{inter}>$ of the coupling network, the effective number of support edges $M$, and the size $s$ of the effective communities. Fig.~\ref{fig5} further illustrates the relationship between the critical threshold point $p_c^A$ and these four parameters. It can be observed that when $<k_{inter}>$ is relatively large, the critical threshold $p_c^A$ decreases as the coupling network's average degree increases, as shown in Fig.~\ref{fig5}(a). This implies that an increase in the number of support edges effectively enhances the system's robustness. However, when $<k_{inter}>$ is relatively small, $p_c^A$ remains at 1, indicating that the initial network is already at the collapse boundary. This is due to the constraints on the effective support edge number $M$ in the effective nodes, and the network only gains stability when $<k_{inter}>$ exceeds $M$. Similarly, when $M$ takes larger values, different networks under different average degrees also exhibit $p_c^A=1$, as shown in Fig.~\ref{fig5}(b). As the average degree increases, $p_c^A$ decreases. However, this decrease is not limitless, as with further increases in the average degree, the critical point gradually stabilizes. From Fig.~\ref{fig5}(c), it is evident that when M is less than $<k_{inter}>$, $p_c^A$ shows an approximately linear relationship with M, which means that as the conditions of effective edges are gradually strengthened, $p_c^A$ also proportionally increases. However, when the value of $M$ approaches or exceeds $<k_{inter}>$, the initial network collapses. Another factor influencing the conditions of effective nodes is the community size $s$. As seen inFig.~\ref{fig5}(d), when $s$ is small, the conditions for effective nodes are relatively relaxed, and the critical threshold point is smaller. As $s$ increases, the size of the community to which effective nodes belong also increases, and at this point, $p_c^A$ becomes larger and gradually approaches the critical point of the model under the maximum connected component, as indicated by the dashed line in the figure.
    
\begin{figure}[ht]
    \centering

        \includegraphics[width=3.3in]{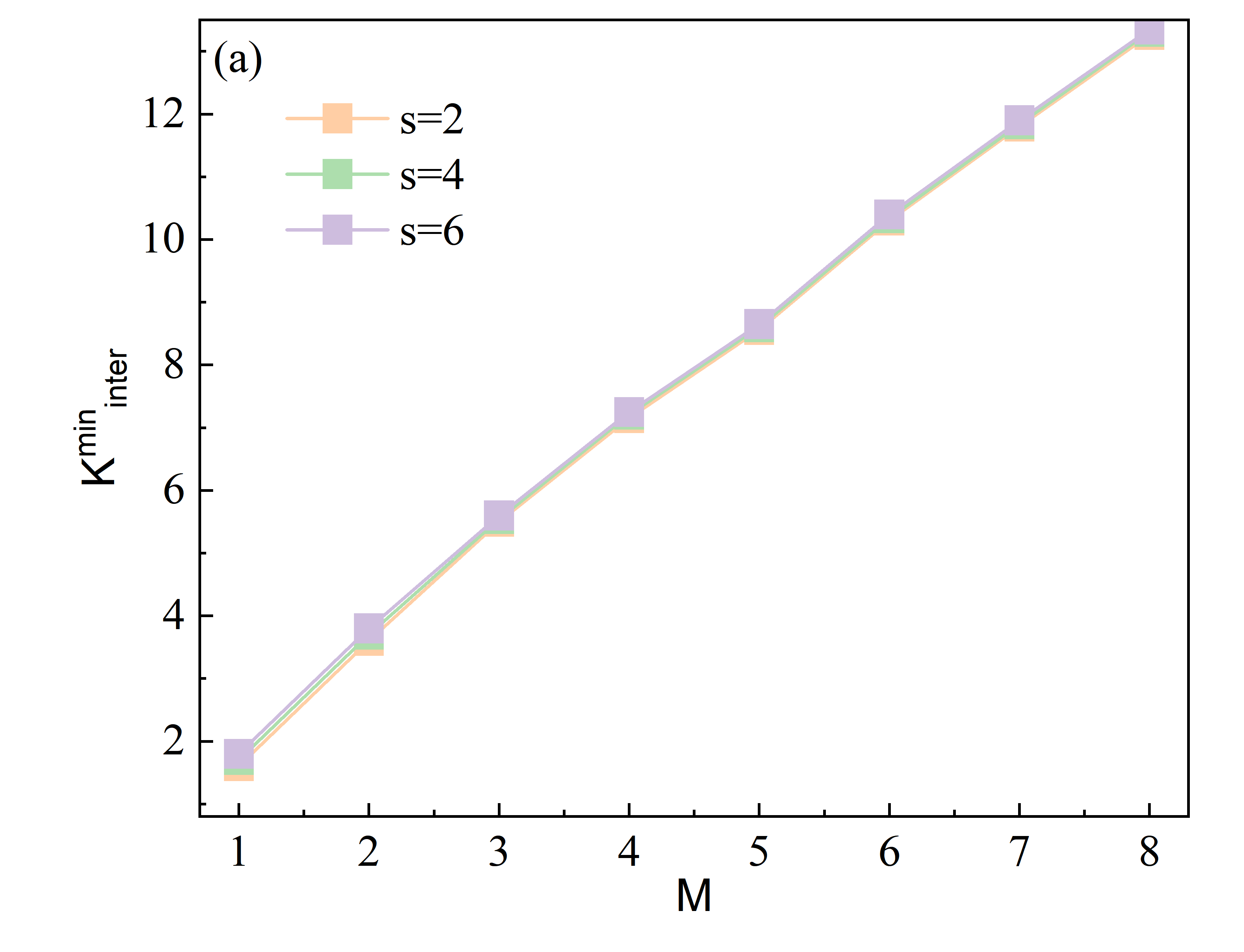}
        \includegraphics[width=3.3in]{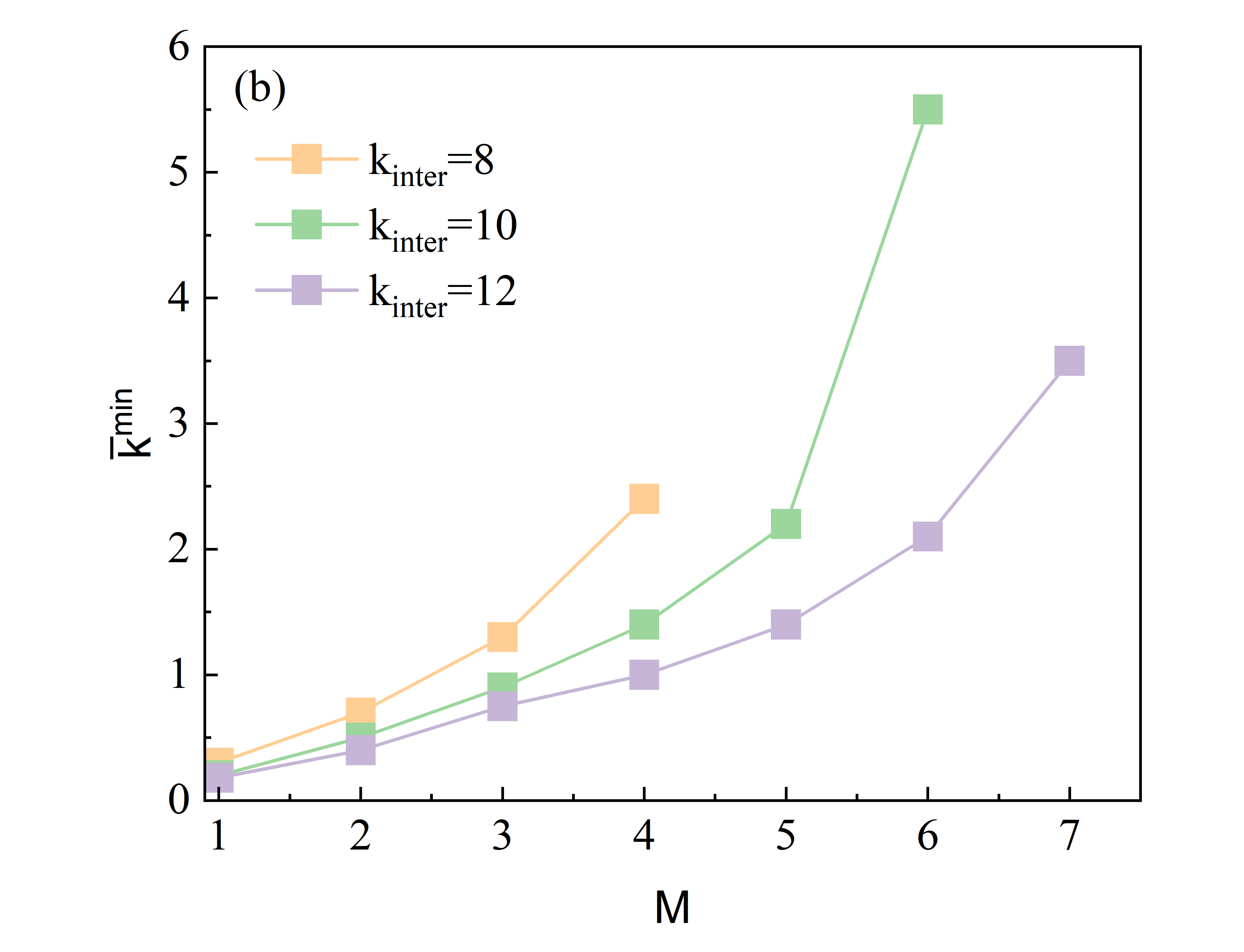}
       
    \hspace{2mm}\vspace{0mm}
    \begin{spacing}{1.5}
    \captionsetup{font=small}
    \caption{
    \fontsize{12}{16}\selectfont
    Minimal $k_{inter}$ and $k$ as a function of $M$ to avoid collapse of the initial coupled ER network. (a) $<k>=4$, (b) $s=4$. }
    \label{fig6}
    \end{spacing}
\end{figure}

    From Fig.~\ref{fig5}(a, b), it can be observed that there are parameters for which $p_c^A=1$, indicating that the network is already at the collapse boundary when subjected to an attack. Here, the parameters maintaining the initial robustness of the network, denoted as $<k_{inter}>$, are respectively represented as $<k_{inter}^{min}>$ and $<k_{min}>$. Fig.~\ref{fig6}(a) illustrates the variation of $<k_{inter}^{min}>$ with changes in the number of effective support edges $M$ under different community sizes $s$. It can be seen that with the growth of $M$, the required minimum number of effective support edges also increases proportionally. Since smaller values of community size s result in more effective nodes, their demand for $<k_{inter}>$ is weaker compared to larger values of $s$, as shown in Fig.~\ref{fig6}(a). In Fig.~\ref{fig6}(b), when $s=4$, the variation of $<k_{min}>$ with $M$ is depicted. With the increase in $M$, the minimum average degree required to maintain the basic stability of the system substantially rises. Additionally, the density of the coupling network can effectively alleviate the requirements for the internal average degree of the network. When the coupling average degree $<k_{inter}>$ is relatively large, the numerical value of $<k_{min}>$ is significantly lower.

\subsection{4.2 Scale-free network robustness analysis after cascade failure}
     The exploration of networks characterized by a power-law degree distribution introduces a fresh lens through which we can comprehend and dissect intricate systems. The attributes and dynamics exhibited by networks conforming to a power-law distribution bear profound significance, furnishing us with tools to model, dissect, and refine real-world systems.  

    The SF network adheres to a power-law degree distribution, and its generating function can be expressed as $G_0(x)=\sum_{k=0}^{\infty}{{\frac{(k+1)^{1-\lambda}-k^{1-\lambda}}{(R+1)^{1-\lambda}-{r}^{1-\lambda}}}x^k}\vspace{1.5ex}$. In this expression, $r$ and $R$ denote the minimum and maximum degree values within the SF network, correspondingly. The parameter $\lambda$ embodies the power-law exponent, while $k$ signifies the range of degrees. $P(k) = \frac{(k+1)^{1-\lambda} - k^{1-\lambda}}{(R+1)^{1-\lambda} - r^{1-\lambda}}$ is the degree distribution.
  \begin{figure}[ht]
    \centering
    
        \includegraphics[width=3in]{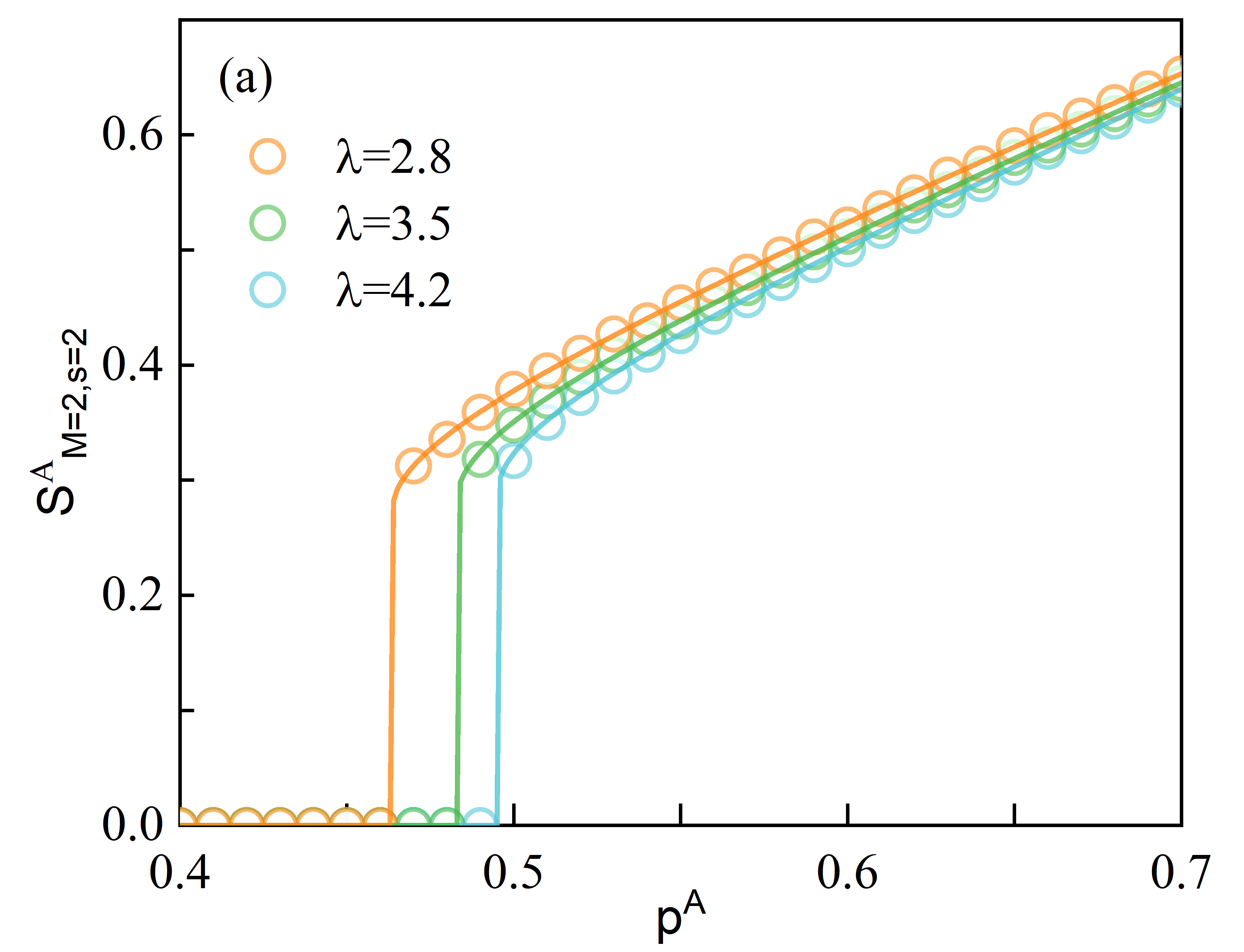}
        \includegraphics[width=3in]{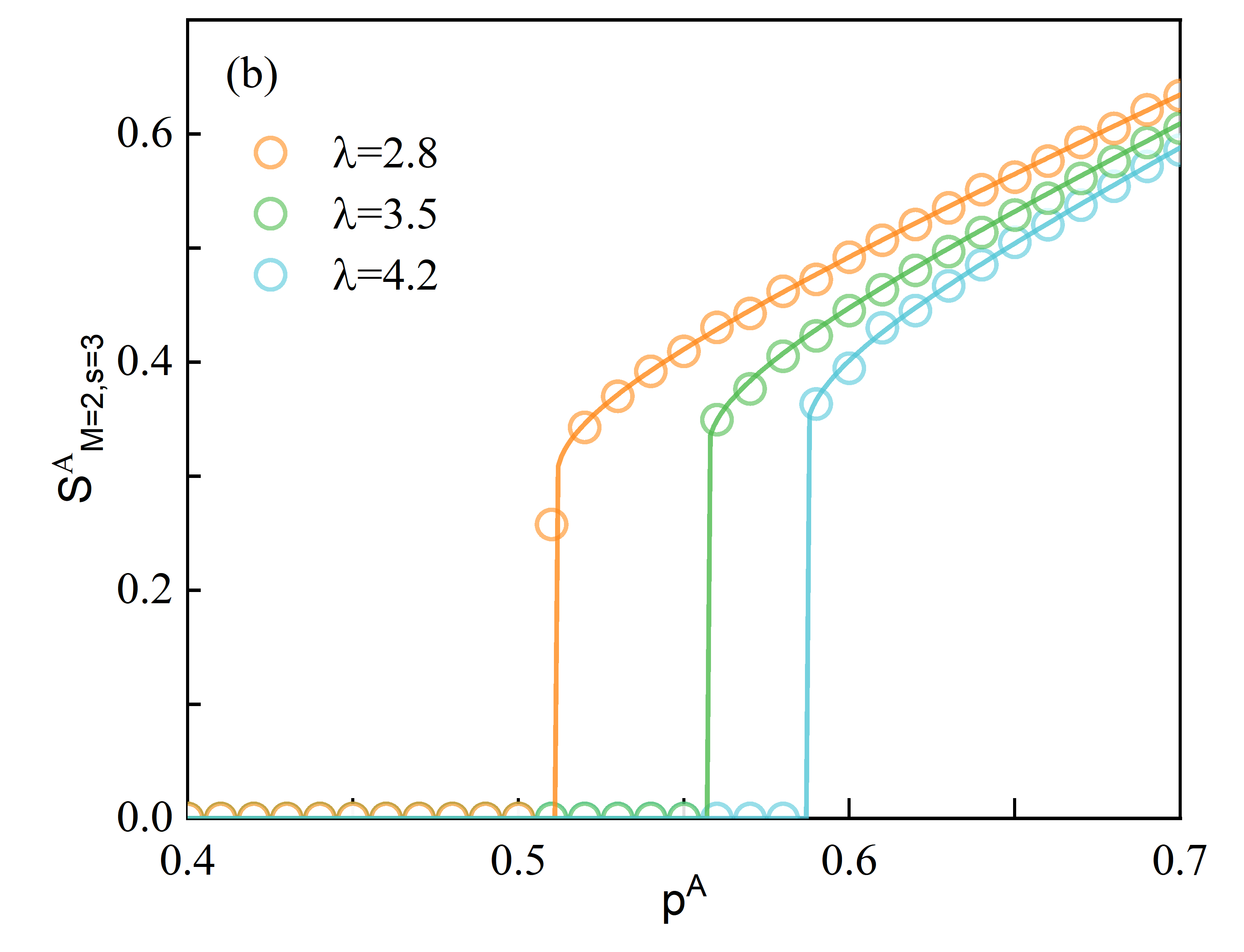}
        \includegraphics[width=3in]{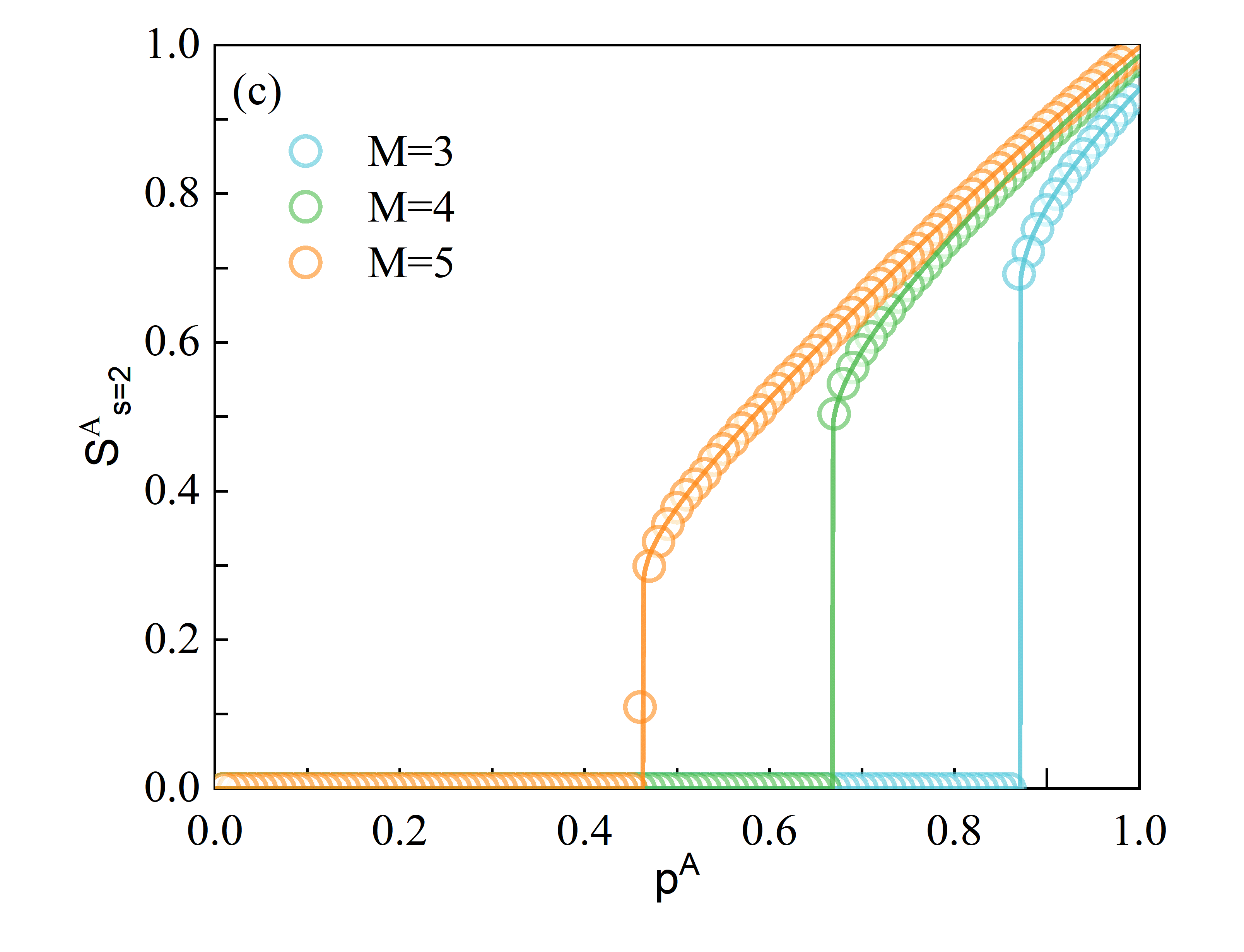}
        \includegraphics[width=3in]{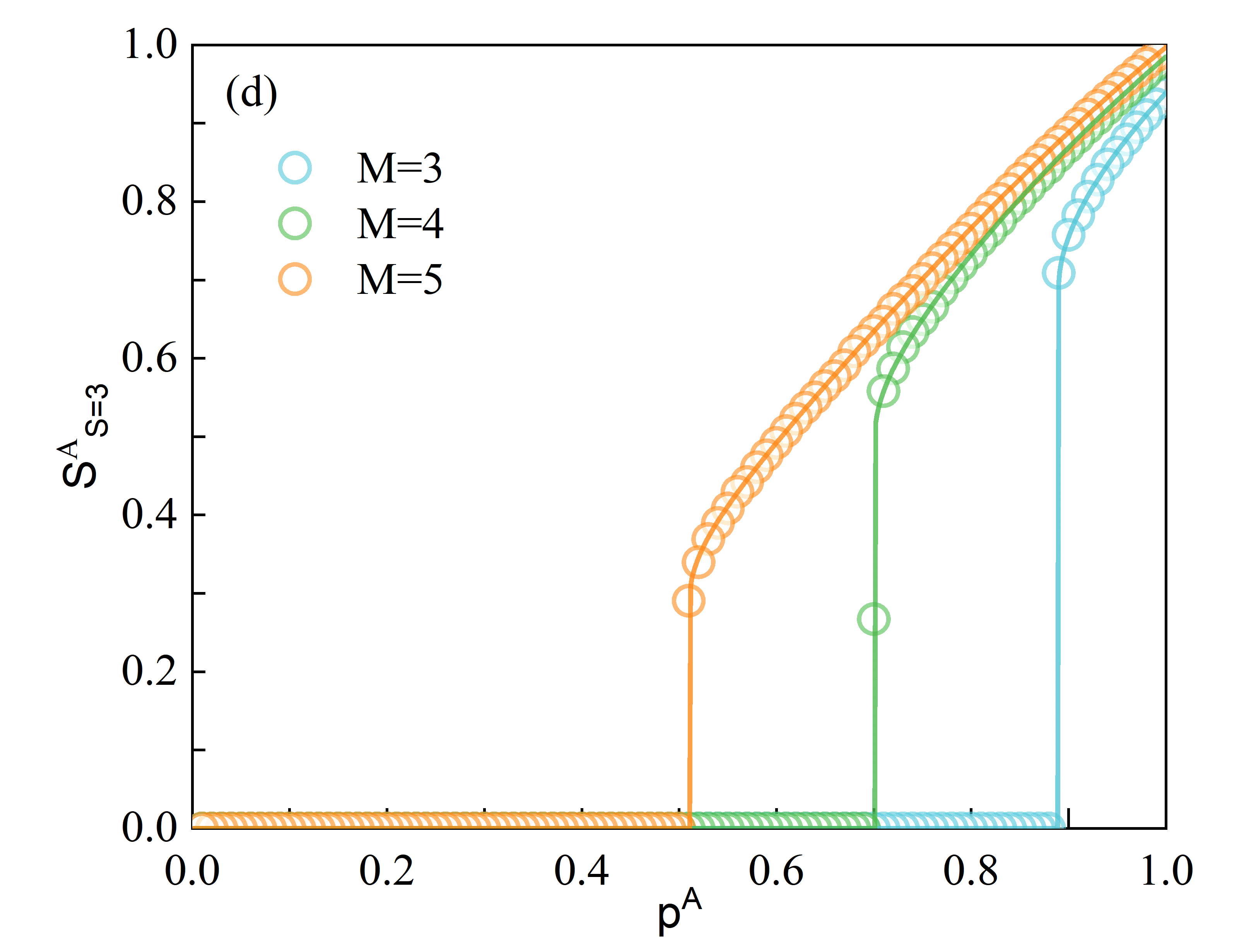}
        
    \hspace{2mm}\vspace{0mm}
    \begin{spacing}{1.5}
    \captionsetup{font=small}
    \caption{
    \fontsize{12}{16}\selectfont
    $S^A_{M,s}$ as a function of $p^A$ for different parameters of $\lambda$, $M$, $s$, and $K_{inter}$: (a) $M=2$, $s=2$,${k_{inter}}^A={k_{inter}}^B=k_{inter}=12$; (b)$M=4$, $s=2$,${k_{inter}}^A={k_{inter}}^B=k_{inter}=12$; (c)  $s=2$, ${k_{inter}}^A={k_{inter}}^B=k_{inter}=8$ and $\lambda=2.8$; (d) $s=3$, ${k_{inter}}^A={k_{inter}}^B=k_{inter}=8$ and $\lambda=2.8$. Circles represent simulation results and solid lines represent theoretical analysis results.}
    \label{fig7}
    \end{spacing}
\end{figure}
   The generating function of the power-law network and the generating function of its transcendence degree distribution are as follows:
\begin{equation}
\begin{cases}
G_0(x)=\sum_{k=0}^{\infty}{{\frac{(k+1)^{1-\lambda}-k^{1-\lambda}}{(R+1)^{1-\lambda}-{r}^{1-\lambda}}}x^k}\vspace{1.5ex},
\\

H(x)=\frac{\sum_{k=0}^{\infty}{{\frac{(k+1)^{1-\lambda}-k^{1-\lambda}}{(R+1)^{1-\lambda}-{r}^{1-\lambda}}}kx^{k-1}}}{\sum_{k=0}^{\infty}{{\frac{(k+1)^{1-\lambda}-k^{1-\lambda}}{(R+1)^{1-\lambda}-{r}^{1-\lambda}}}k}}.

\end{cases}
\label{eq15}
\end{equation}
  From Fig.~\ref{fig7}(a), it can be observed that, in the SF-coupled network model, the analytical results (lines) are in complete agreement with the simulations (symbols).  Furthermore, as the fraction of remaining nodes, $p^A$, increases, Network $A$ exhibits a first-order phase transition in its largest connected component. By increasing the number of support edges under the condition of effective nodes, it can be observed that the critical point of Network $A$ also increases, as shown in Fig.~\ref{fig7}(b). Additionally, the critical threshold point $p_c^A$ gradually decreases with an decrease in the coupling network's power-law exponent $\lambda$.   Fig.~\ref{fig7}(c) and Fig.~\ref{fig7}(d) illustrates the influence of the size $s$ of the effective connected component on the network. A smaller value of $s$ implies that there are more nodes in communities with sizes greater than or equal to $s$. Hence, the critical point $p_c^A$ increases with an increase in the value of $s$.
 
\begin{figure}[ht]
    \centering
        \includegraphics[width=3in]{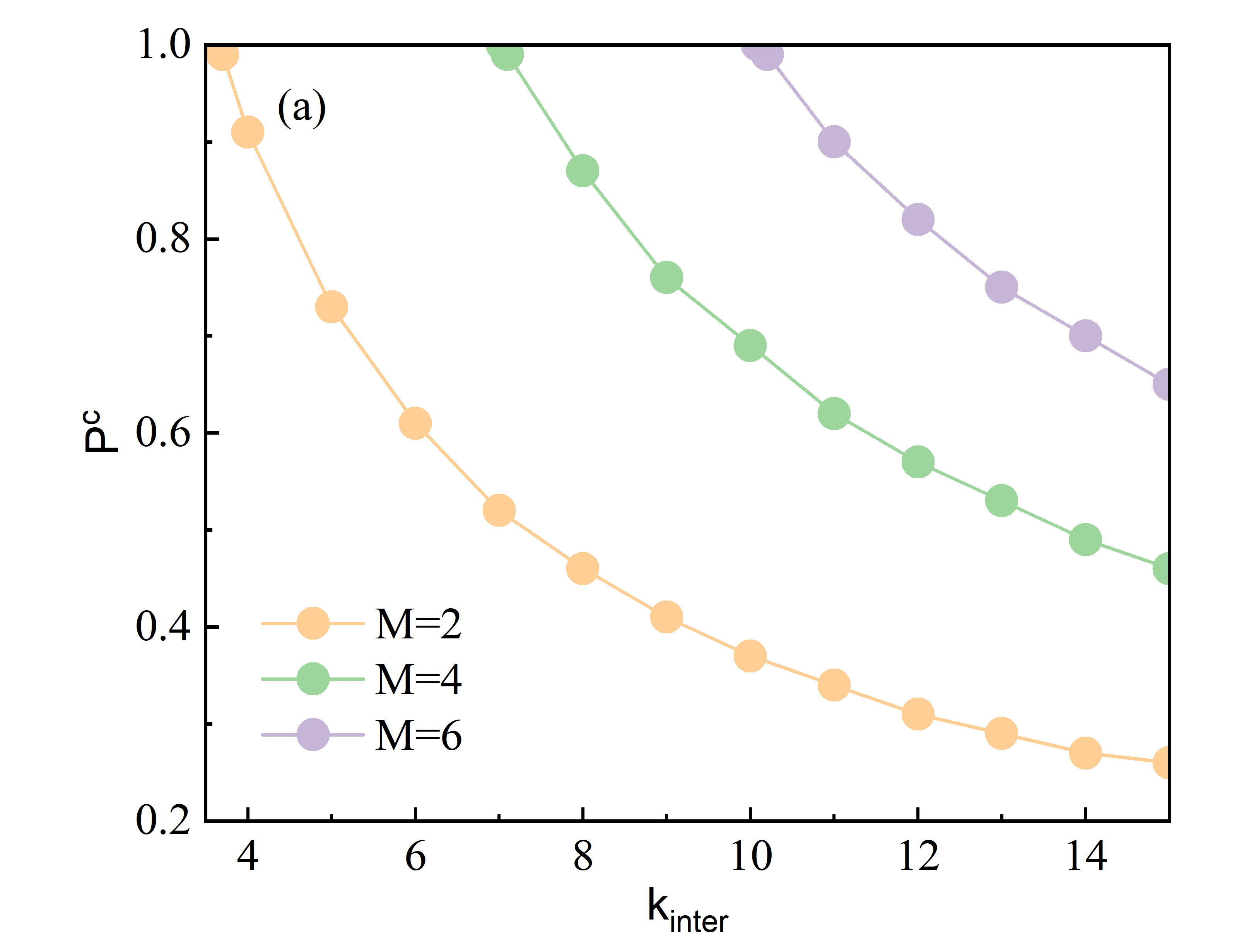}
        \includegraphics[width=3in]{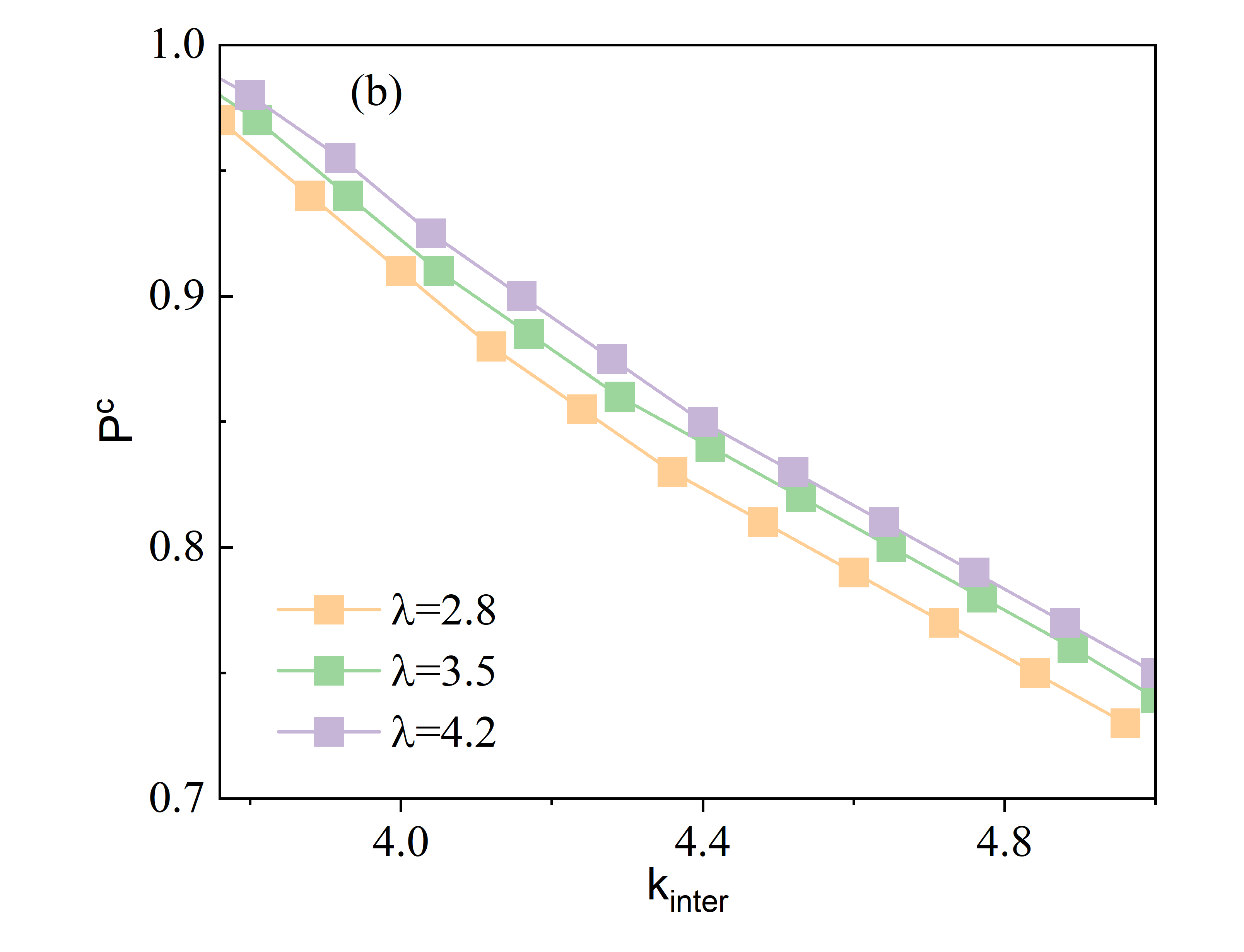}
        \includegraphics[width=3in]{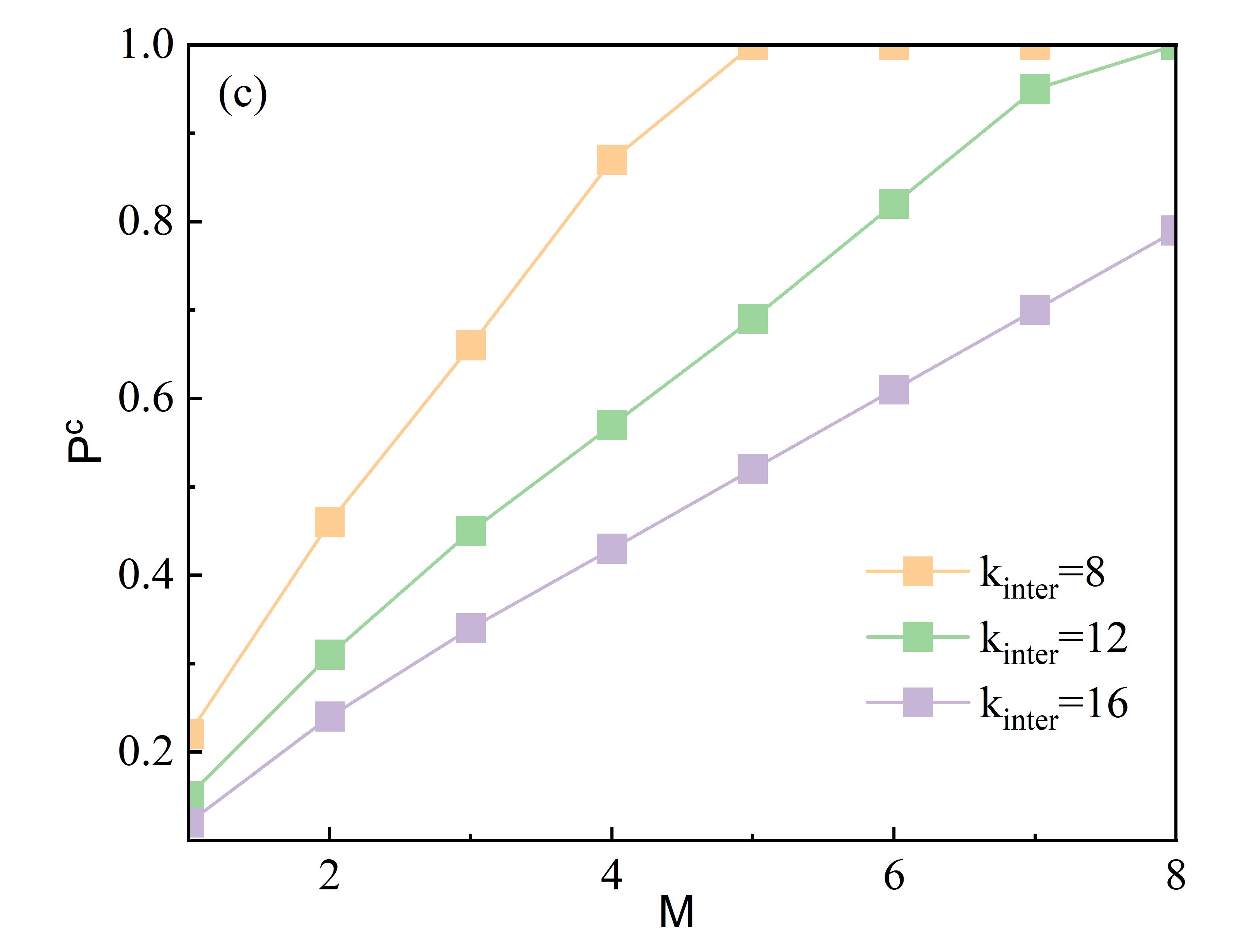}
        \includegraphics[width=3in]{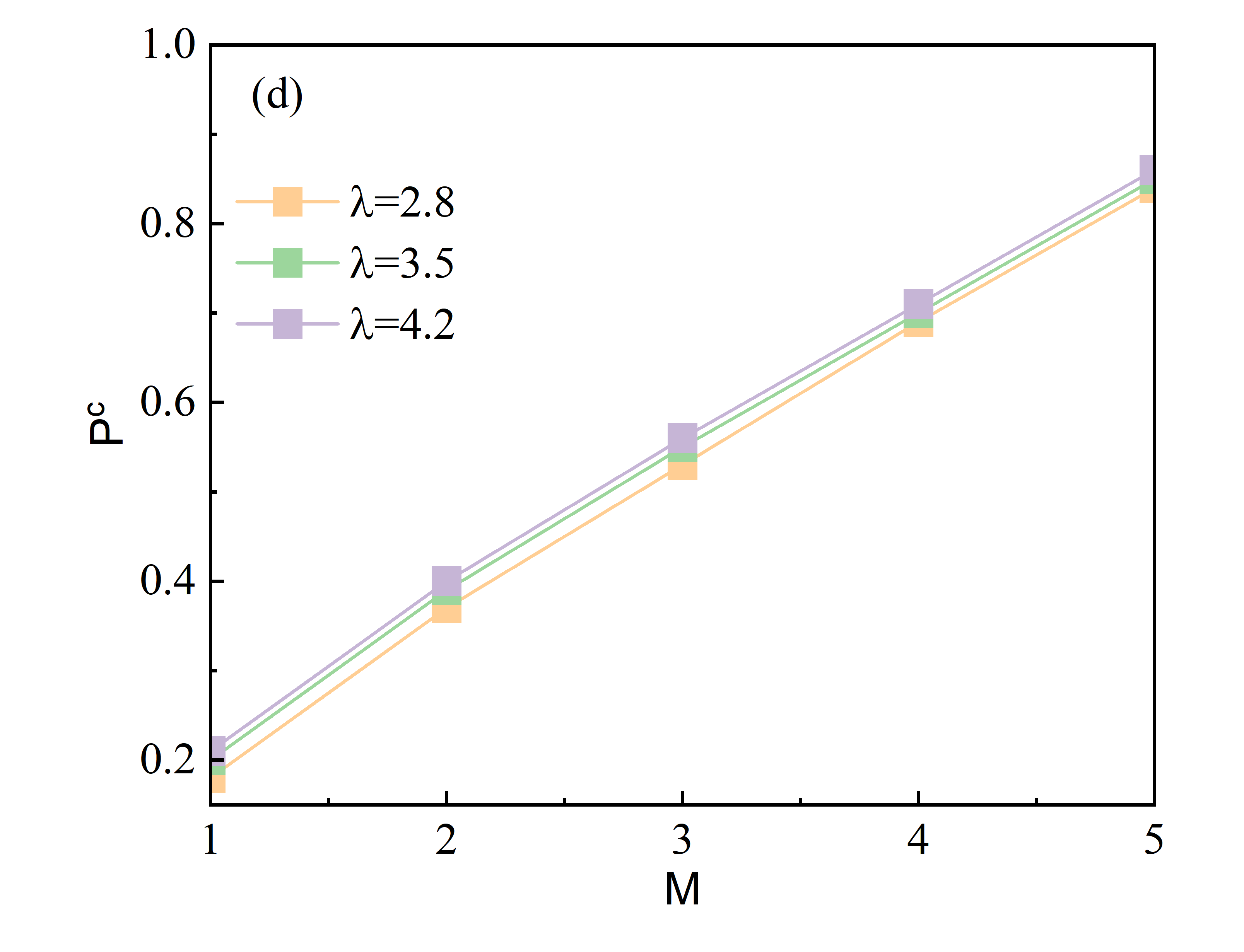}
        
    \hspace{2mm}\vspace{0mm}
    \begin{spacing}{1.5}
    \captionsetup{font=small}
    \caption{
    \fontsize{12}{16}\selectfont
    Critical points ${p_A}^c$ as a function of different parameters. (a) $s=2$ and $\lambda=2.8$. (b) $s=2$ and $M=2$. (c)$s=2$ and $\lambda=2.8$. (d) $\lambda=2.8$, $k_{inter}= 12$.} 
    \label{fig8}
    \end{spacing}
\end{figure}
   Fig.~\ref{fig7} both indicate that the primary factors affecting this coupled network are the power-law exponent $\lambda$ of the sub-network, the average degree $<k_{inter}>$ of the coupling network, the effective number of support edges $M$, and the size $s$ of the effective communities. Fig.~\ref{fig8} further illustrates the relationship between the critical threshold point $p_c^A$ and these four parameters. It can be observed that when $<k_{inter}>$ is relatively large, the critical threshold $p_c^A$ decreases as the coupling network's average degree increases, as shown in Fig.~\ref{fig8}(a). This implies that an increase in the number of support edges effectively enhances the system's robustness. This is due to the constraints on the effective support edge number $M$ in the effective nodes, and the network only gains stability when $<k_{inter}>$ exceeds $M$. As the power-law exponent decreases, $p_c^A$ decreases as shown in Fig.~\ref{fig8}(b). From Fig.~\ref{fig8}(c) and Fig.~\ref{fig8}(d), it is evident that when $M$ is less than $<k_{inter}>$, $p_c^A$ shows an approximately linear relationship with $M$, which means that as the conditions of $<k_{inter}>$ and $\lambda$ are gradually strengthened, $p_c^A$ also proportionally increases.

\section{5. Conclusion}
 \fontsize{12}{16}\selectfont
   This study introduces a novel model that introduces the concepts of a finite cluster size $s$ and multiple effective dependency links $M$ to simulate the evolutionary behavior of a coupling system under attack, aimed at enhancing system robustness. This implies that the connectivity of the coupling system necessitates functional nodes within local networks of clusters equal to or exceeding $s$, along with a minimum of $M$ external support links to withstand risks and failures effectively. This approach closely aligns with real-world scenarios. The theoretical analysis of the model is corroborated through simulations involving coupled Poisson and power-law networks. After cascade failures, a first-order phase transition emerges in the system, culminating in stabilization. Furthermore, the study reveals that augmenting the interconnection density between the Cyber-Phsical systems becomes imperative to uphold network robustness, particularly when $s$ and $M$ are substantial. Moreover, the research delves into establishing the minimal connection density requisite for the survival of a coupled network encompassing $s$ clusters and $M$ effective support links, thereby ensuring the coupling system's continuity. To sum up, the study's findings contribute to a deeper comprehension of coupled complex systems in the real world, offering insights for designing robust infrastructures and efficacious risk mitigation strategies.
\section{CRediT authorship contribution statement}

\section{Acknowledgment}
This research is supported by grants from the National Natural Science Foundation of China (Grant No. 62373169, 61973143, 71690242), National Statistical Science Research Project (Grant No. 2022LZ03), Special Project of Emergency Management Institute of Jiangsu University (Grant No. KY-A-08), the National Key Research and Development Program of China (Grant No. 2020YFA0608601), and the Jiangsu Postgraduate Research and Innovation Plan (Grant No. KYCX22$\_$3601).

\bibliographystyle{unsrtnat}

\bibliography{ref}

\end{document}